\documentclass[12pt]{article}
\usepackage{amsfonts,amsthm,amsmath,amssymb,upgreek}
\usepackage[paper=letterpaper,margin=1in]{geometry}
\usepackage[export]{adjustbox}
\usepackage[numbers,sort&compress]{natbib}
\usepackage{color}
\usepackage{graphicx}% Include figure files
\usepackage{dcolumn}% Align table columns on decimal point
\usepackage{bm}% bold math
\usepackage{hyperref}% add hypertext capabilities
\usepackage{float}
\usepackage{makecell}
\usepackage[caption=false]{subfig}
\usepackage{hyperref}
\hypersetup{
    colorlinks=true,
    linktoc=page,    %set to all if you want both sections and subsections linked
    linkcolor=blue,
    urlcolor=magenta
}
\usepackage{tikz}
\usetikzlibrary{decorations.pathmorphing}
\usepackage{overpic}

\newcommand{\be}{\begin{equation}}
\newcommand{\ee}{\end{equation}}
\newcommand{\bea}{\begin{eqnarray}}
\newcommand{\eea}{\end{eqnarray}}
\newcommand{\dd}{\text{d}}

\newcommand{\e}{\varepsilon}
\newcommand{\bmF}{\bar{\mathcal{F}}}

\newcommand{\R}{\mathcal{R}}
\newcommand{\daggerfootnote}[1]{%
    \renewcommand{\thefootnote}{\fnsymbol{footnote}}%
    \footnote[2]{#1}
    \renewcommand{\thefootnote}{\arabic{footnote}}%
}
\begin{document}
%===============================

\thispagestyle{empty}

\vspace*{.5cm}
\begin{center}

{\bf {\Large On the Universal Cuspy Behavior in Black Hole Shadows}\\
\vspace{1cm}}

 {\bf Peng Cheng$^{a}$\daggerfootnote{p.cheng.nl@outlook.com} and Si-Jiang Yang$^{b,c,d}$}\\
  \bigskip \rm
  
\bigskip

 a) Center for Joint Quantum Studies and Department of Physics, \\School of Science, Tianjin University, Tianjin 300350, China\\
b) Center for Theoretical Physics, School of Physics and Optoelectronic Engineering, Hainan University, Haikou 570228, China\\
c) Lanzhou Center for Theoretical Physics, Key Laboratory of Theoretical Physics of Gansu Province,
Key Laboratory of Quantum Theory and Applications of MoE,
Gansu Provincial Research Center for Basic Disciplines of Quantum Physics, Lanzhou University, Lanzhou 730000, China\\
d) Institute of Theoretical Physics $\&$ Research Center of Gravitation,
School of Physical Science and Technology, Lanzhou University, Lanzhou 730000, China

\rm

\vspace{1.5cm}
{\bf Abstract}
\end{center}
\begin{quotation}

This work investigates the universality of cusp formation in the shadows of compact objects. The emergence of cusps is accompanied by three interrelated phenomena: a topological charge transition, an equal-area law governing the self-intersecting structure, and universal critical scaling behavior. We demonstrate that, because these phenomena originate from the global morphology of the shadow, they are fundamentally independent of specific spacetime metric details and apply across diverse models. These features are systematically analyzed for the Kerr black hole endowed with a running Newton coupling. By extending our framework to rotating traversable wormholes, we confirm that the same universal behavior persists in more general compact objects. Our study uncovers the universality underlying cusp formation, offering a model-independent framework for characterizing possible non-Kerr shadow morphologies.
\end{quotation}

\vspace{1cm}

\setcounter{page}{0}
\setcounter{tocdepth}{2}
\setcounter{footnote}{0}

%%%%%%%%%%%%%%%%%%%%%%%%%%%%%%%%%%%%%%%%%%%%%%%%%%%%%%%%%%%%%%%%%%%%%%%%%%%%%%%%%%%%%%%%%%%%%%%%%%%%
% MAIN BODY
%%%%%%%%%%%%%%%%%%%%%%%%%%%%%%%%%%%%%%%%%%%%%%%%%%%%%%%%%%%%%%%%%%%%%%%%%%%%%%%%%%%%%%%%%%%%%%%%%%%%

\newpage
{\noindent} \rule[-10pt]{16.5cm}{0.05em}\\
\tableofcontents
{\noindent} \rule[-10pt]{16.5cm}{0.05em}\\
%\pagebreak

%%%%%%%%%%%%%%%%%%%%%%%%%%%%%%%%%%%%%%%%%%%%%%%%%%%%%%%%%%%%%%%%%%%%%%%%%%%%%%%%%%%%%%%%%%%%%%%%%%%%

\section{Introduction}
\label{intro}

The optical appearance and shadow of a compact gravitating object, fundamentally rooted in the phenomena of gravitational lensing and photon ring formations, have long served as a cornerstone of strong-field gravity phenomenology~\cite{Synge:1966,Luminet:1979}. Crucially, following the Event Horizon Telescope's groundbreaking release of the M87* and Sagittarius A* images~\cite{EventHorizonTelescope:2019ggy,EventHorizonTelescope:2022wxl,EventHorizonTelescope:2021dqv}, the study of black hole shadows has evolved into a highly mature field~\cite{Perlick:2021aok,Wei:2013kza,Wang:2018prk}. It now offers profound insights into extracting intrinsic black hole parameters~\cite{Hioki:2009na,Bambi:2019tjh,Liu:2025wwq,Tsukamoto:2014tja,Tsukamoto:2017fxq}, testing the fundamental tenets of general relativity, such as the no-hair theorem~\cite{Johannsen:2010ru,Broderick:2013rlq} and tightly constraining a myriad of modified gravity theories~\cite{Vagnozzi:2022moj}. Consequently, shadow analyzes have been broadly extended to exotic compact objects, such as naked singularities~\cite{Shaikh:2018lcc,Khodadi:2024ubi}, gravastars~\cite{Rosa:2024bqv,Sakai:2014ypa}, and wormholes~\cite{Cunha:2018uzx,Nedkova:2013msa,Cheng:2026wyk,Bronnikov:2021liv}. 

% The optical appearance and shadow of a compact gravitating object, fundamentally rooted in the phenomena of gravitational lensing and photn ring formations~\cite{Gralla:2019xty,Cunha:2018acu}, have long served as a cornerstone of strong-field gravity phenomenology~\cite{Synge:1966,Luminet:1979}. Crucially, following the Event Horizon Telescope's groundbreaking release of the M87* and Sagittarius A* images~\cite{EventHorizonTelescope:2019ggy,EventHorizonTelescope:2022wxl,EventHorizonTelescope:2021dqv}, the study of black hole shadows has evolved into a highly mature field. It now offers profound insights into extracting intrinsic black hole parameters~\cite{Hioki:2009na}, testing the fundamental tenets of general relativity, such as the no-hair theorem~\cite{Johannsen:2010ru} and tightly constraining a myriad of modified gravity theories~\cite{}~\cite{Gralla:2019xty,Cunha:2018acu,Perlick:2021aok}. Consequently, shadow analyzes have been broadly extended to exotic compact objects, such as naked singularities~\cite{}, gravastars~\cite{}, and wormholes~\cite{Wei:2013kza,Sakai:2014ypa,Cunha:2018uzx,Nedkova:2013msa,Wang:2018prk,Cheng:2026wyk}. 

Among the myriad of observable signatures, such as polarization patterns~\cite{Chen:2020qyp,Hou:2024qqo,Chen:2024jkm} and intricate photon ring sub-structures~\cite{Wang:2025hzu,Gan:2021xdl,Johnson:2019ljv,Himwich:2020msm}, the exact morphology of the shadow boundary remains of paramount importance. It intrinsically encodes the global phase-space structure of photon orbits within strong gravitational regimes, offering a direct, high-precision probe into the underlying spacetime geometry~\cite{Gralla:2019xty,Cunha:2018acu}.
% Among various observable signatures, the morphology of the shadow boundary is of paramount importance. It intrinsically encodes the global structure of photon orbits within strong gravitational regimes, offering a direct probe into the underlying spacetime geometry.
%Among various potentially observable features, the morphology of the shadow boundary is of paramount importance, as it encodes the global structure of photon orbits in strong gravitational fields and directly reflects the geometric properties of the background spacetime. 
An especially compelling feature of this morphology is the formation of cusps, which are sharp turning points along the boundary that signal a departure from smoothness~\cite{Cunha:2017,Qian:2022,Wang:2016paq,Chen:2025jay}. Far from being mere numerical artifacts, the appearance of these cusps is a genuine physical signature revealing the fundamental characteristics of photon orbits~\cite{Cunha:2025oeu,Wei:2026zwu}.

The emergence of a cusp in a black hole shadow serves as a striking signature of physics beyond the Kerr paradigm, and is associated with theoretical importance by drawing an analogy between the transition of shadow morphology and thermodynamic systems \cite{Wei:2026zwu}.
It was shown that the shadow morphological change of the rotating Konoplya-Zhidenko (KZ) non-Kerr black hole possesses three fundamental features.
First, the morphological change is related to topological charge flips from $\delta=+1$ to $\delta=-1$.
Second, the transition of morphology exactly follows the equal-area law.
Third, one can identify a critical point marking the onset of cusp formation. 
Near this critical point, the universal scaling behavior is governed by a critical exponent of $1/2$, placing the system within the mean-field universality class. 

While the discovery of a topological phase transition and critical scaling behavior in the shadow of the KZ black hole provides a compelling analogy between black hole optics and statistical physics, a fundamental question remains: are these features unique to this particular Kerr deformation, or do they represent a universal principle governing cusp formation in compact object shadows? To answer this, we must examine physically distinct systems that also exhibit cuspy shadows and, if those features apply across models, uncover their underlying physical drivers. Addressing these issues is the primary objective of this work. By extending our analysis to diverse classes of compact objects, we elucidate the universal mechanisms dictating both cusp formation and the self-intersecting dynamics of the shadow boundary.

Through a detailed study of cusp formation in a Kerr black hole with a running Newton coupling, $G$, we show that this phenomenon is consistently accompanied by three universal features: a topological charge transition, an equal-area law, and a universal critical exponent.
Moreover, we demonstrate that these features are not restricted to specific geometry details but governed by universal geometric and topological principles. We further verify this universality by examining rotating traversable wormholes, which exhibit the identical topological transition, equal-area law, and square-root critical scaling near the onset of cusp formation. Ultimately, this universality places the formation of cuspy shadows within a model-independent framework, revealing a profound connection between black hole optics, global topology, and critical phenomena.
%The universality is further checked by examining rotating traversable wormholes, which exhibit the same topological transition, equal-area law, and square-root critical scaling near the onset of cusp formation. 
%This universality places cuspy shadow formation in a model-independent framework and reveals a deep connection between black hole optics, global topology, and critical phenomena. 
%The universality of the cusp formation may provide fundamental non-Kerr features in high-resolution shadow observations.

This paper is organized as follows. 
In Section \ref{KZ}, we briefly review the cuspy shadow and universal features for the KZ non-Kerr black hole. 
Section \ref{RG-Kerr} presents a detailed analysis of the Kerr black hole with a running Newton coupling, demonstrating that the same three universal features emerge when the spin parameter crosses a critical threshold. 
In Section \ref{universal}, we uncover the geometric and topological mechanisms underlying this universality and further test its validity in a rotating traversable wormhole, confirming that cusp universality applies to more general scenarios. 
Finally, Section \ref{con} summarizes our main results and discusses future issues.

%%%%%%%%%%%%%%%%%%%%%%%%%%%%%%%%%%%%%%%%%%%%%%%%%%%%%%%%%%%%%%%%%%%%%%%%%%%%%%%%%%%%%%%%%%%%%%%%%%%%

\section{Cuspy shadow for the KZ non-Kerr black hole}
\label{KZ}

We review the framework for studying the cuspy behavior of the KZ black hole, following~\cite{Wei:2026zwu}.
We will briefly introduce the shadow cast by the rotating KZ black hole \cite{Wang:2017hjl}, which exhibits a cuspy structure.
The cuspy structure was shown to exhibit striking signatures \cite{Wei:2026zwu}, particularly the topological phase transition, the gravitational equal-area law, and critical phenomena in the cuspy shadow.

\subsubsection*{The shadow of KZ rotating non-Kerr black hole}

The KZ metric modifies the Kerr black hole by introducing a deformation term, which modifies the horizon structure, ergoregion, and singularity behavior. %This metric can be regarded as a low-energy effective description of possible black hole solutions in quantum gravity models.
The study of shadows cast by such black holes provides a useful theoretical arena for characterizing possible deviations from Kerr shadow morphology.
%The study of shadows cast by such black holes provides intuitive and testable predictions for next-generation black hole observations.
The metric in Boyer-Lindquist coordinates is given by:
\begin{equation}
	ds^2 = -\frac{\Delta}{\Sigma} \left( \dd t - a \sin^2 \theta \, \dd \phi \right)^2 + \frac{\Sigma}{\Delta} \, \dd r^2 + \Sigma \, \dd \theta^2 + \frac{\sin^2 \theta}{\Sigma} \left[ a \, \dd t - (r^2 + a^2) \, \dd \phi \right]^2,
\end{equation}
%\begin{equation}
%\begin{aligned}
%ds^2 = &-\left(1 - \frac{2GMr^2 + \e}{r\Sigma}\right)dt^2 + \frac{\Sigma}{\Delta} dr^2 + \Sigma d\theta^2 + \sin^2\theta \left[r^2 + a^2 + \frac{(2GMr^2 + \e)a^2\sin^2\theta}{r\Sigma}\right]d\phi^2 \\
%&-\frac{2(2GMr^2 + \e)a\sin^2\theta}{r\Sigma} dt d\phi,
%\end{aligned}
%\end{equation}
with
\begin{equation}
\Delta = a^2 + r^2 - 2GMr - \frac{\e}{r}, \quad \Sigma = r^2 + a^2\cos^2\theta.
\end{equation}
The metric is described by three parameters: the mass $M$, the rotating parameter $a$, and the deformation parameter $\e$\footnote{Note that we use $\e$ to represent the deformation parameter, rather than $\eta$ used in \cite{Konoplya:2016pmh,Konoplya:2016jvv,Wang:2016paq,Wang:2017hjl,Wei:2026zwu}, to avoid notation conflict with the impact parameters. }. It reduces to the normal Kerr metric when $\e\to 0$.

\begin{figure}[hbt]
    \centering
    \subfloat[$\e>\e_c$]{\includegraphics[width=0.28\textwidth]{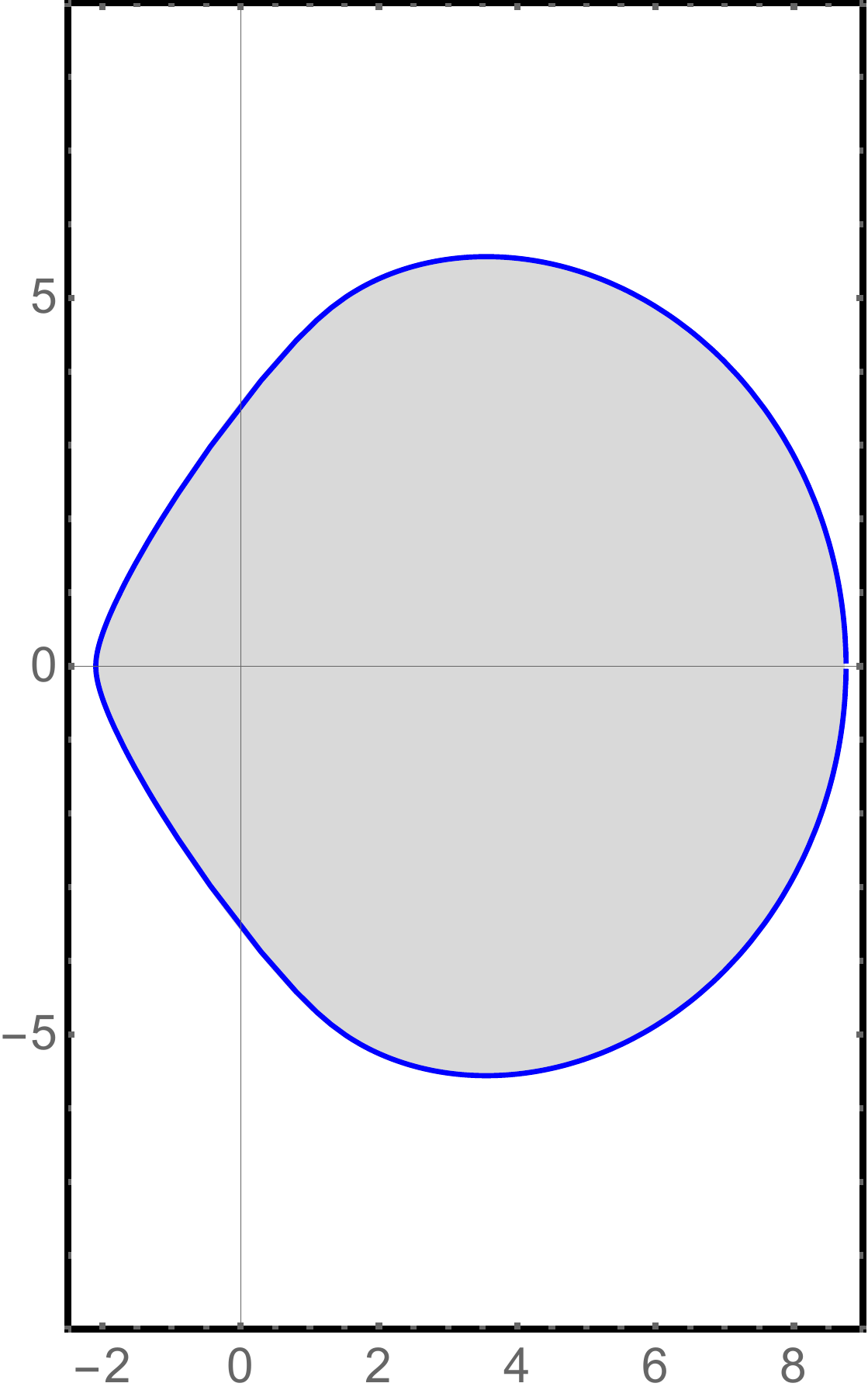}}
    \subfloat[$\e=\e_c$]{\includegraphics[width=0.28\textwidth]{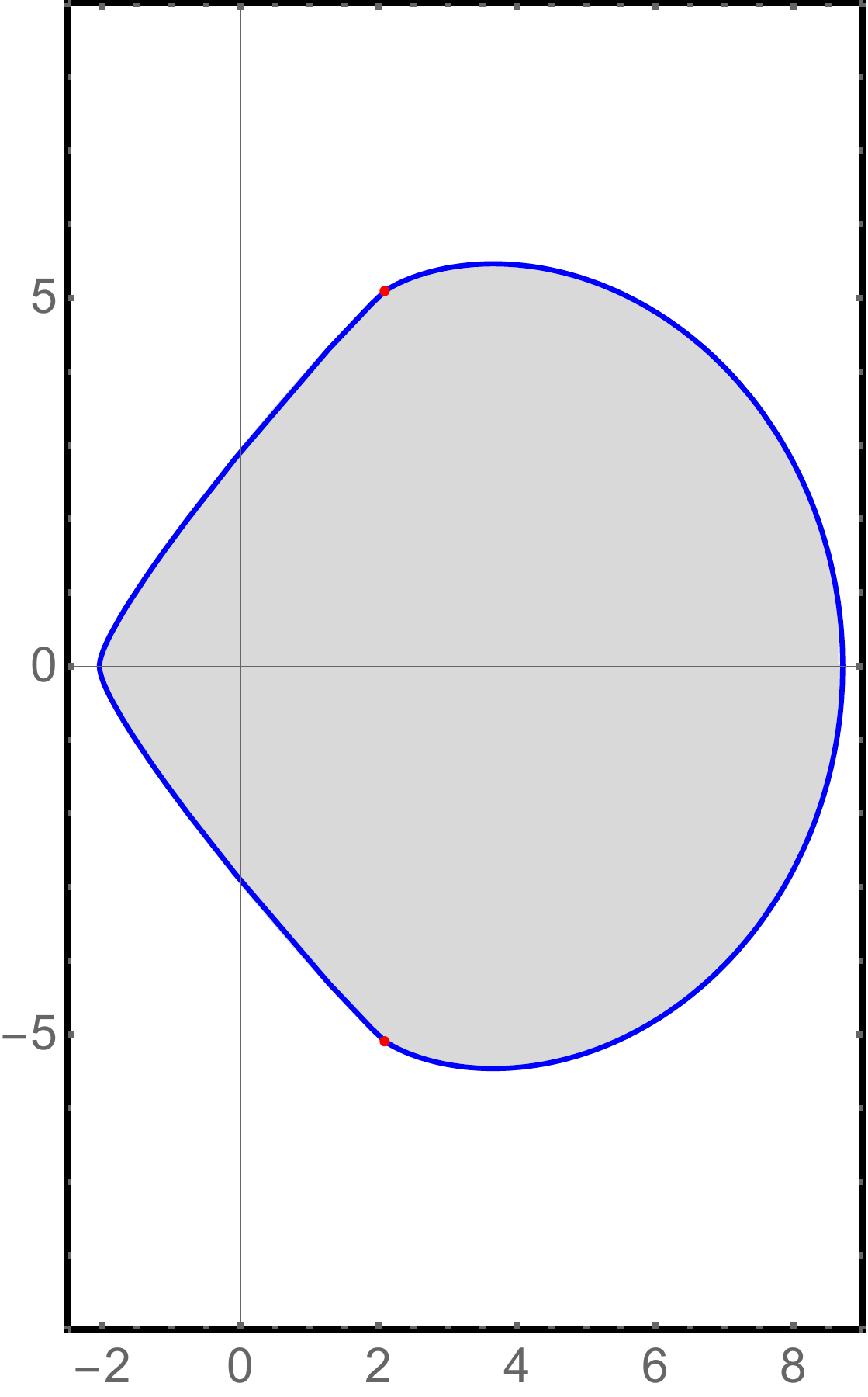}}
    \subfloat[$\e<\e_c$\label{fig:KZ_3}]{\includegraphics[width=0.28\textwidth]{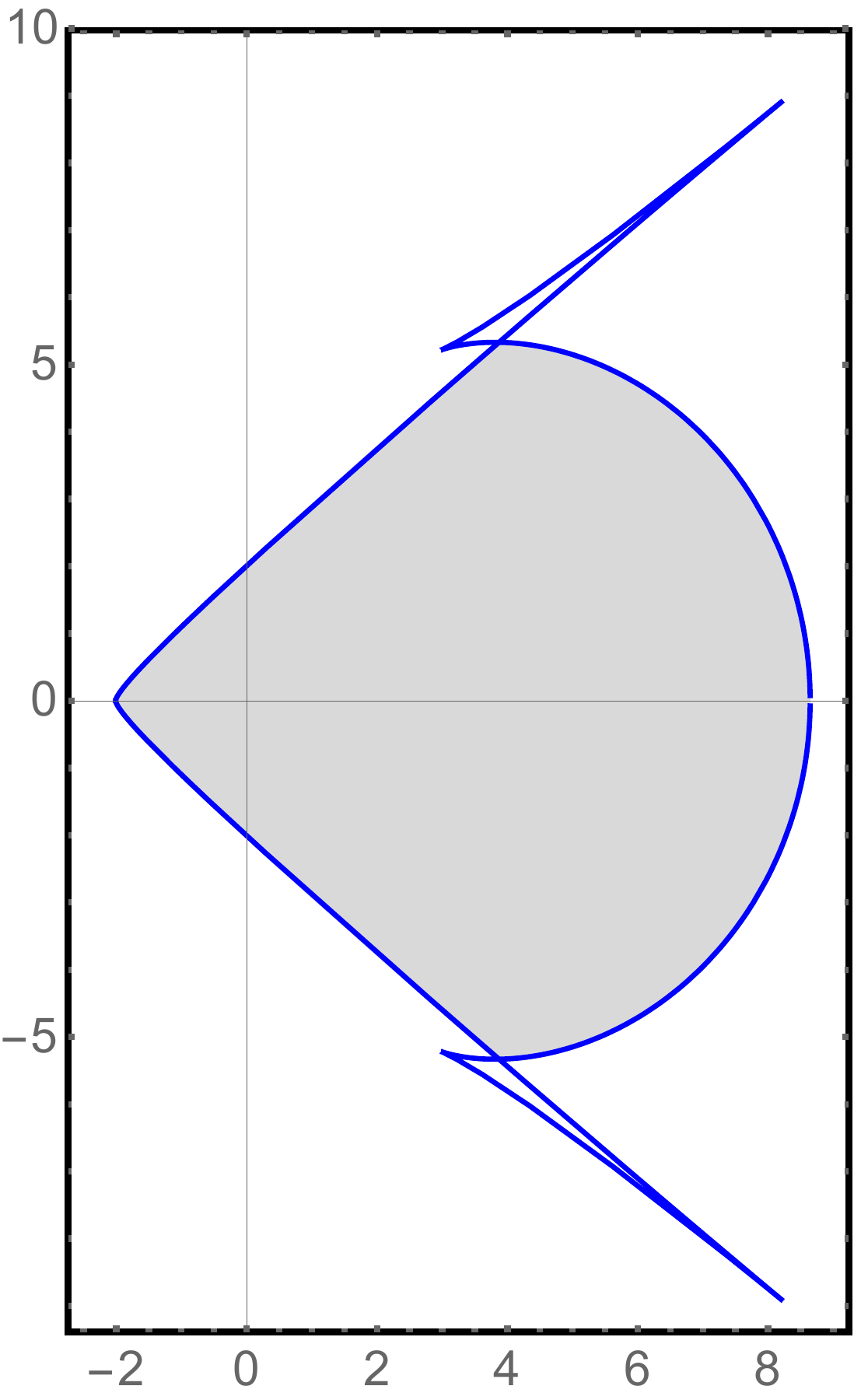}}
    \subfloat{\begin{tikzpicture}
    \draw [thick,->] (0,-1)--(0,1) node[above=5pt]{$\beta$};
    \draw [thick,->] (0,0)--(1,0) node[midway,below=3pt]{$\alpha$};
    \node at (0,-3.65) {};
    \end{tikzpicture}}
    \caption{Black hole shadows for the KZ black hole with $a_* = 2$. For deformation parameter $\e>\e_c$, the shadow is smooth, while the shadow develops a cuspy structure for $\e<\e_c$. The critical case is shown in (b).}
    \label{fig:KZ_1_3}
\end{figure}
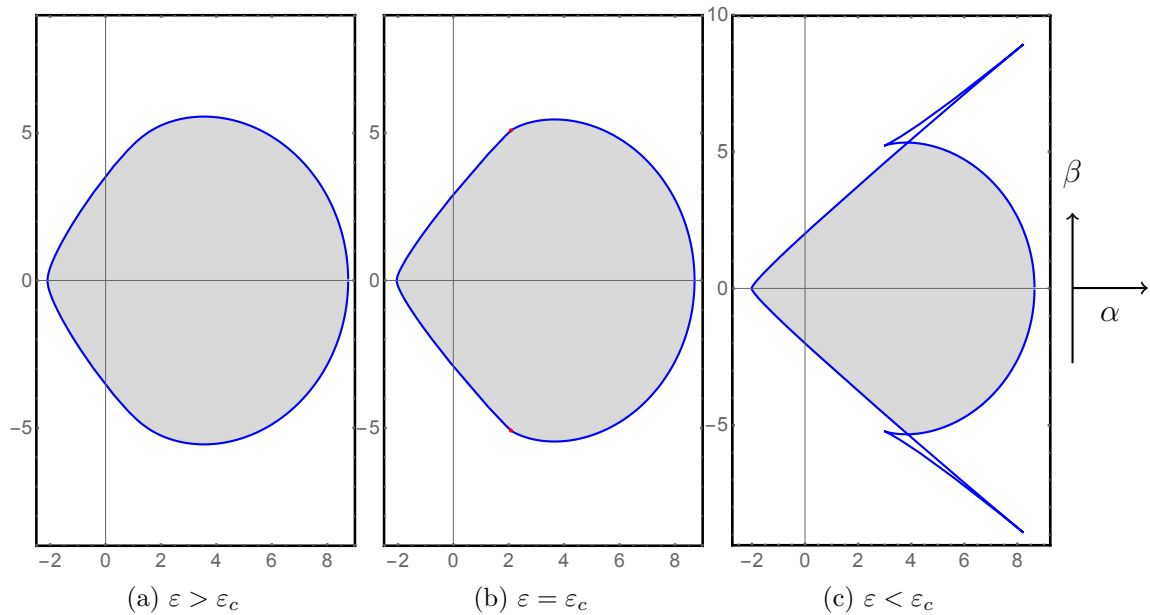

%Due to stationarity and axisymmetry, there are two conserved quantities for the geodesic particles: the energy $E = -p_t$ and the angular momentum $L = p_\phi$.
Through the Hamilton-Jacobi method, one can show that the unstable circular photon orbits can be expressed by Bardeen's impact parameters $(\alpha,\beta)$. 
For an observer at the equatorial plane ($\theta_0=\pi/2$), the orbits in coordinates $(\alpha, \beta)$ can be explicitly expressed as~\cite{Wang:2017hjl}
\begin{eqnarray}
\begin{split}
	\alpha &= \frac{6GMr^4 - 2r^5 + 5\e r^2 + a^2(\e - 2r^2(GM + r))}{2a(GM - r)r^2 - a\e},\\
	\beta &=\pm \frac{r^2\sqrt{8a^2(2GMr^3 + 3r\e) - (6GMr^2 - 2r^3 + 5\e)^2}}{a(2r^2(r - GM) + \e)}.
\end{split}\label{eq:ab1}
\end{eqnarray}
The slope of the unstable circular orbits can be expressed as
\begin{equation}
	\mathcal{F}=\frac{\dd\beta}{\dd \alpha}\,.
\end{equation}
Due to the presence of the deformation parameter $\e$, an event horizon is preserved for super-spinning regimes.
With fixed spin at $a_*=a/(GM) = 2$, one can depict how the shadow of a KZ black hole changes as the deformation parameter $\e$ varies in Fig. \ref{fig:KZ_1_3}. Note that in the figure, $\alpha$ and $\beta$ are redefined to be dimensionless using parameters $G$ and $M$.
As illustrated in the figure, three distinct regimes are clearly displayed. For $\e> \e_c$ (where $\e_c$ denotes the critical value of the deformation parameter), the shadow exhibits a smooth, quasi-circular (or D-shaped) contour similar to that of a standard Kerr black hole. When $\e$ approaches $\e_c=1.052(GM)^3$ from above, the shadow begins to show a structural transition marked by a noticeable corner: a precursor to the more dramatic change that follows. For $\e<\e_c$, the shadow develops a cuspy structure along its boundary. 

%\subsection{Universal physics behind the cuspy shadow}
%\label{sec:UniKZ}

The shadows of KZ non-Kerr black holes exhibit distinct cuspy behaviors, marking a transition from typical quasi-circular contours to cuspy geometries. 
As demonstrated in~\cite{Wei:2026zwu}, a systematic investigation reveals that these features are dictated by universal principles governing the physics near the critical point. 
This phenomenology is underpinned by a unified theoretical framework that elegantly synthesizes topological phase transitions, the equal-area law, and critical phenomena.

\subsubsection*{Topological phase transition}

The family of unstable spherical photon orbits, when projected onto the observer’s celestial plane, defines the shadow boundary: a closed contour whose global structure can be rigorously characterized by its topological charge.
The transition from a standard quasi-circular contour to a swallowtail behavior, as illustrated in Fig.~\ref{fig:KZ_1_3}, can be effectively explored through the evolution of this topological invariant.

Evaluated via the Gauss-Bonnet theorem, the topological charge $\delta$ of the shadow, defined as the winding number of its boundary ~\cite{Wei:2018xks}, undergoes a sign inversion, signaling a topological phase transition. This change in the topological charge of the unstable circular orbits is driven by the deformation parameter $\varepsilon$ of the KZ geometry. Specifically, for $\varepsilon > \varepsilon_c$, the shadow boundary remains a smooth, simple closed curve yielding a topological charge of $\delta = +1$. This consequently places it in the exact same topological class as the canonical Kerr black hole shadow.
However, as $\e$ is reduced below $\e_c$, a pair of swallowtails emerges on the unstable circular orbits.
Consequently, the total topological charge transfers to $\delta = -1$ for the shadow boundary with swallowtail. 
The shift is not a continuous deformation but signifies a fundamental topological phase transition. 
The topological charge distinguishes cuspy shadows from smooth ones by placing them into distinct topological classes, a phenomenon analogous to the discrete ``topological number flip'' observed in condensed matter systems.
%The topological charge distinguishing cuspy shadows from smooth ones as belonging to separate topological classes, analogous to a discrete ``topological number flip'' observed in condensed matter systems. 
%This establishes that the formation of a cusp is a critical, topology-altering event in gravitational lensing.

\subsubsection*{Equal-area law}

%\begin{figure}[H]
%\begin{center}
%  \subfloat[$\e<\e_c$\label{fig:KZ-area1}]{\includegraphics[width=0.45\textwidth]{KZ_area1.pdf}}~~~~
%  \subfloat[$\e=\e_c$\label{fig:KZ-area2}]{\includegraphics[width=0.45\textwidth]{KZ_area2.pdf}}
%  \caption{Gravitational equal-area law for the cuspy black hole shadow. For deformation parameter $\e/M^3=0.7$ smaller than the critical value, there is an equal-area law; whereas at the critical point we have an inflection point with $\mathcal{F}_1=\mathcal{F}_2$.}\label{fig:KZ-area}
%\end{center}
%\end{figure}

The self-intersecting boundary of the shadow shown in Fig.~\ref{fig:KZ_1_3} is reminiscent of the free energy swallowtail behavior in liquid-gas systems~\cite{Kubiznak:2012wp}.
%Similar to the thermodynamic system, there indeed is an equal-area law governing the cuspy shadow.
The local slope of the shadow curve for the KZ black hole exhibits a characteristic non-monotonic behavior, and the equal-area law is satisfied. 
This allows the identification of a self-intersection point $(\alpha_i, \beta_i)$ where two distinct photon orbits coincide on the celestial plane. 
For the deformation parameter smaller than the critical value $\e<\e_c$, it can be easily verified numerically that in the $(\alpha,\mathcal{F})$ diagram, the two areas enclosed between the curve and a vertical line at $\alpha_i$ are equal. 
At the critical point $\e=\e_c$, we have an inflection point.
Above the critical value, the slope of the shadow is monotonic, and we have the smooth quasi-circular black hole shadow. 
The equal-area law provides a precise method to locate the self-intersection point for the black hole shadow.
 
 %, thereby establishing the equal-area law as a robust and predictive tool for analyzing shadow morphology in non-Kerr spacetimes. Moreover, it builds a relation between the geometry of a cuspy black hole shadow and the equal-area law.

\subsubsection*{Critical phenomena and critical exponent}

%\begin{figure}[H]
%\begin{center}
%  \subfloat[Critical exponent $\zeta_1$]{\includegraphics[width=0.45\textwidth]{KZ_ce1.pdf}}\quad
%  \subfloat[Critical exponent $\zeta_2$]{\includegraphics[width=0.45\textwidth]{KZ_ce2.pdf}}
%  \caption{Critical behavior of the KZ black hole with $\alpha_c=2$. The critical exponents are $\zeta_1=\zeta_2=1/2$. 
%  The numerical results near the criticality (represented by red dots) exhibit a scaling behavior consistent with the critical exponents, as validated by their close proximity to the reference blue lines of slope $1/2$.}\label{fig:KZ-ce}
%\end{center}
%\end{figure}

The onset of cusp formation is characterized by universal scaling behavior. 
The critical point, which marks the exact threshold of cusp formation, is geometrically identified as an inflection point in the $(\alpha, \mathcal{F})$ plane.
For the specific case of a black hole with spin $a_* = 2$, this critical point occurs at the deformation parameter $\e_c \approx 1.052(GM)^3$ and the celestial coordinate $\alpha_c \approx 2.088GM$. 
To quantify the behavior near this critical point, one can define the separation between the two intersecting branches of the shadow boundary $\Delta \mathcal{F} = \mathcal{F}_2 - \mathcal{F}_1$ as the order parameter.
This parameter exhibits characteristic power-law scalings with the distance from the critical point, following 
\begin{equation}
	\begin{split}
		\Delta \mathcal{F} &\sim |\alpha - \alpha_c|^{\zeta_1}\,,\\
		\Delta \mathcal{F} &\sim | \e-\e_c |^{\zeta_2}\,.
	\end{split}
\end{equation}
It was proven that the critical exponents robustly converge to the same value
\begin{equation}
	\zeta_1 = \zeta_2 = \frac{1}{2}\,.
\end{equation}
This value of the critical exponent is highly significant, as it places the system within the mean-field universality class. 
%The proof of the universal critical exponent establishes a profound connection between black hole optics and statistical mechanics.

\section{Shadow of running-$G$ Kerr black hole}
\label{RG-Kerr}

The KZ black hole provides a compelling framework for studying cuspy shadow formation, revealing a rich structure of topological and critical phenomena. Nevertheless, a single example, however instructive, is insufficient to establish a general physical principle. To determine the universality of these features, we must investigate alternative physical systems that also exhibit cuspy shadows. A natural candidate for this is the Kerr black hole with a running Newton coupling, a model motivated by quantum-gravity considerations.

In quantum field theory, renormalization group flows imply that coupling constants depend on the energy scale, giving rise to so-called \textit{running couplings}. When quantum effects of gravity are taken into account, the gravitational constant $G$ is also expected to become scale-dependent in quantum gravity frameworks such as asymptotically safe gravity. 
Black holes endowed with a running Newton coupling $G$ can therefore be regarded as effective models of quantum-corrected black hole geometries, providing an effective phenomenological setting in which possible quantum-gravity-inspired corrections to black-hole shadow morphology can be studied.
%providing a phenomenological avenue to test quantum gravity theories through black hole observations.

However, the Newton constant cannot be replaced by an arbitrary running coupling. Instead, its functional form should be constrained by fundamental physical principles. In the context of black holes, the consistency of black hole thermodynamics imposes nontrivial restrictions on the allowed form of the running gravitational coupling $G$. 
The metric of a Kerr black hole with a running Newton constant can be expressed as
\begin{equation}
ds^{2} = -\frac{\Delta}{\Sigma}\left(\dd t - a\sin^{2}\theta \dd \phi\right)^{2} + \frac{\Sigma}{\Delta} \dd r^{2} + \Sigma \dd \theta^{2} + \frac{\sin^{2}\theta}{\Sigma}\left[a \dd t - (r^{2} + a^{2})\dd \phi\right]^{2},
\end{equation}
where
\begin{equation}
\Delta(r) = r^{2} + a^{2} - 2 G(r, M, a) M r, \quad \Sigma = r^{2} + a^{2}\cos^{2}\theta.
\end{equation}
According to the thermodynamic consistency condition, the running Newton constant $G(r, M, a)$ must have a specific form to be consistent with black hole thermodynamics.
Defining dimensionless parameters
\begin{equation}
	r_*\equiv \frac{r}{G_0M},\quad a_*\equiv \frac{a}{G_0M}\,,
\end{equation}
the running Newton constant can be expressed using the following parameterization from the Pad\'{e} by requiring proper ultraviolet and infrared behavior \cite{Chen:2022xjk,Chen:2023wdg}
\begin{equation}
G(z) \equiv G_0 \left( \frac{1 + \sum_{n=1}^\infty e_n z^n}{1 + \sum_{m=1}^\infty f_m z^m} \right), \label{pade}
\end{equation}
where we have
\begin{equation}
	z = \frac{1}{\left( G_0 M^2 \right)^5 \left( r_*^2 + a_*^2 \right) r_*^4},
\end{equation}
and the original Newton constant $G_0$. The parameterization describes a model of nonsingular black holes \cite{Chen:2023wdg}.
In the nonsingular scenario, the black hole can be super-spinning. The running-$G$ black hole case discussed here can be regarded as an explicit example of the situation studied in \cite{Bambi:2008jg}.
In this scenario, $z$ decreases approximately as $1/r_*^6$ with increasing $r_*$. Contributions from higher-order coefficients decay rapidly away from the black hole. Thus, we can retain only the lowest coefficients in \eqref{pade}, neglecting all higher-order terms. In this work, we only focus on the specific case with parameter $f_1=-4$ and neglect all other parameters in \eqref{pade}.
We can work with a given $a_*$ and vary the parameters $e_n$ and $f_n$ to examine the formation of the cuspy shadow; or equivalently, we may fix $e_n$ and $f_n$, and investigate the topological charge of the shadow edge with different values of $a_*$. The latter approach will be adopted in this study.

\begin{figure}[H]
    \centering
    \subfloat[$a_*=1$\label{fig:RGKerr_1}]{\includegraphics[width=0.45\textwidth]{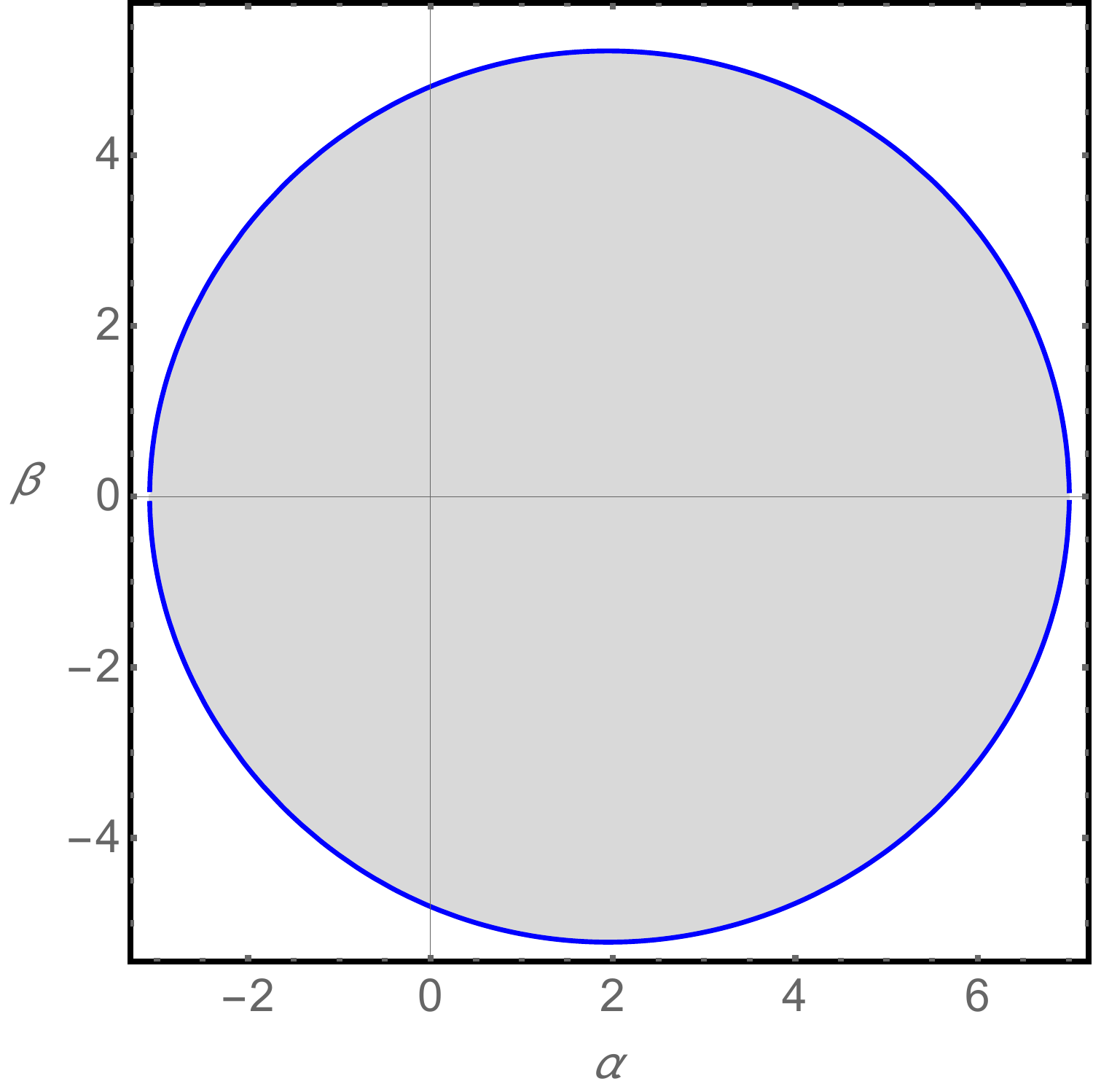}}~~
    \subfloat[$a_*=a_c\approx 2.343$\label{fig:RGKerr_2}]{\includegraphics[width=0.45\textwidth]{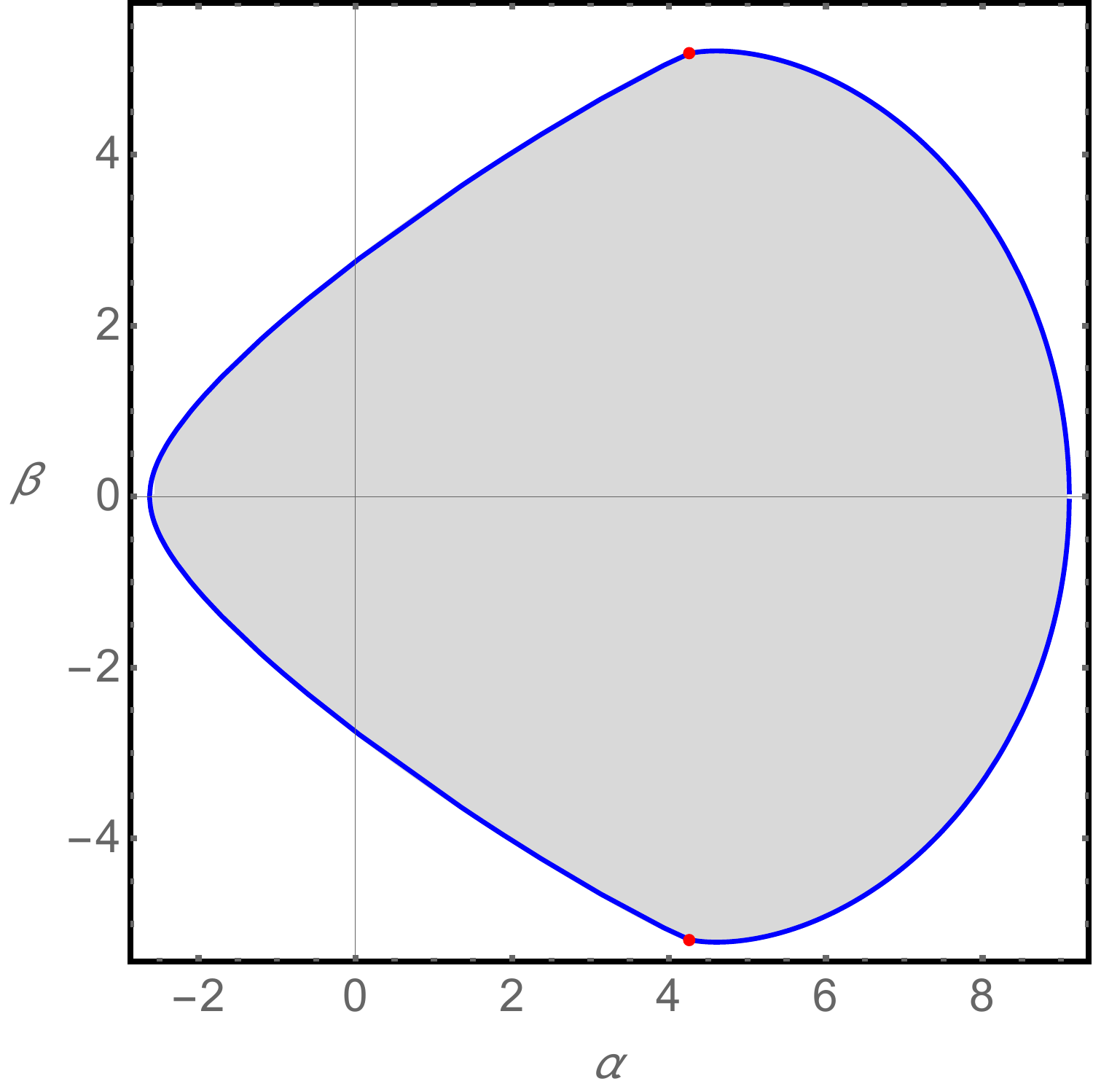}}\\
    \subfloat[$a_*=3$\label{fig:RGKerr_3}]{\includegraphics[width=0.45\textwidth]{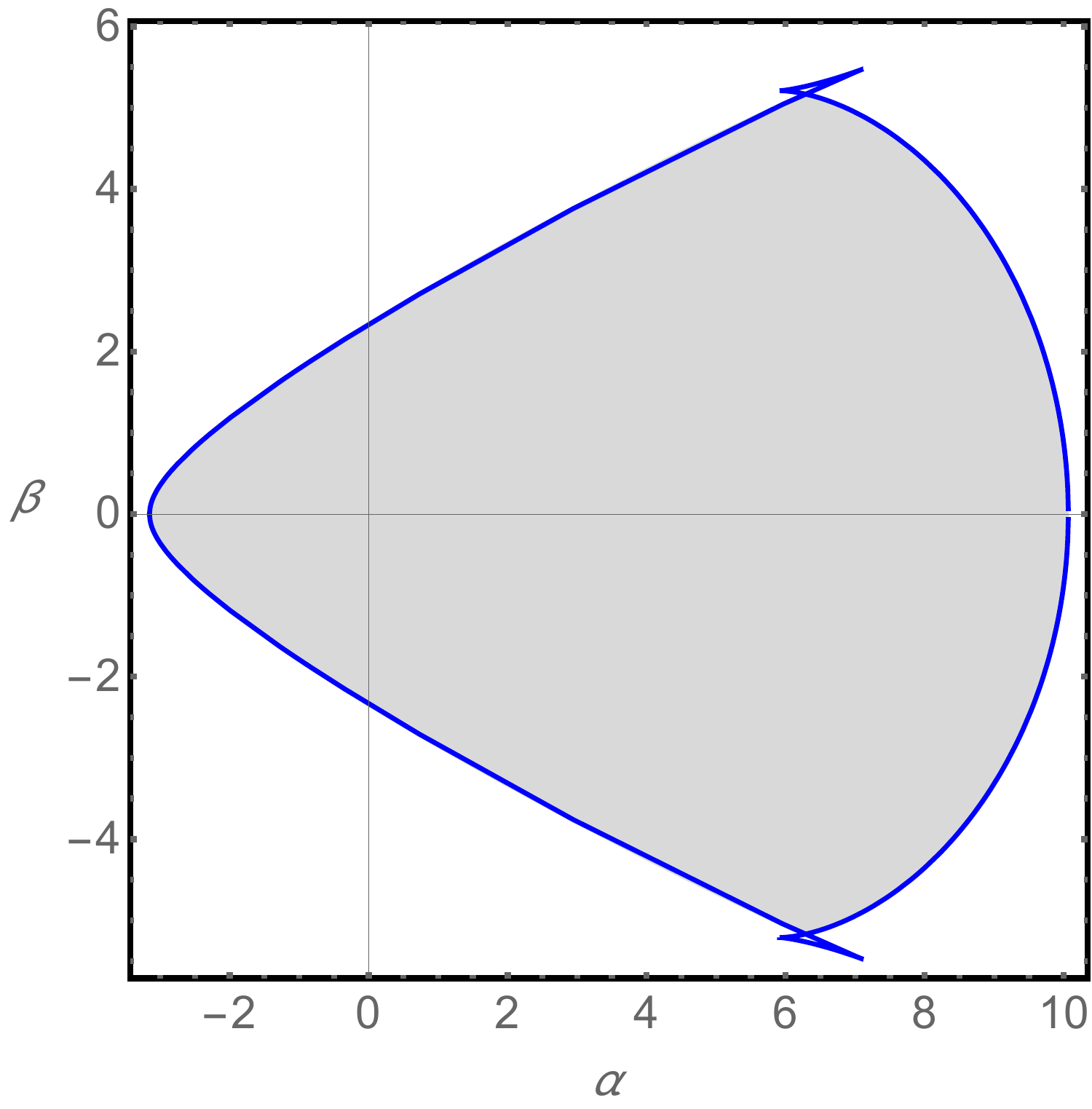}}~~
    \subfloat[$a_*=4$\label{fig:RGKerr_4}]{\includegraphics[width=0.45\textwidth]{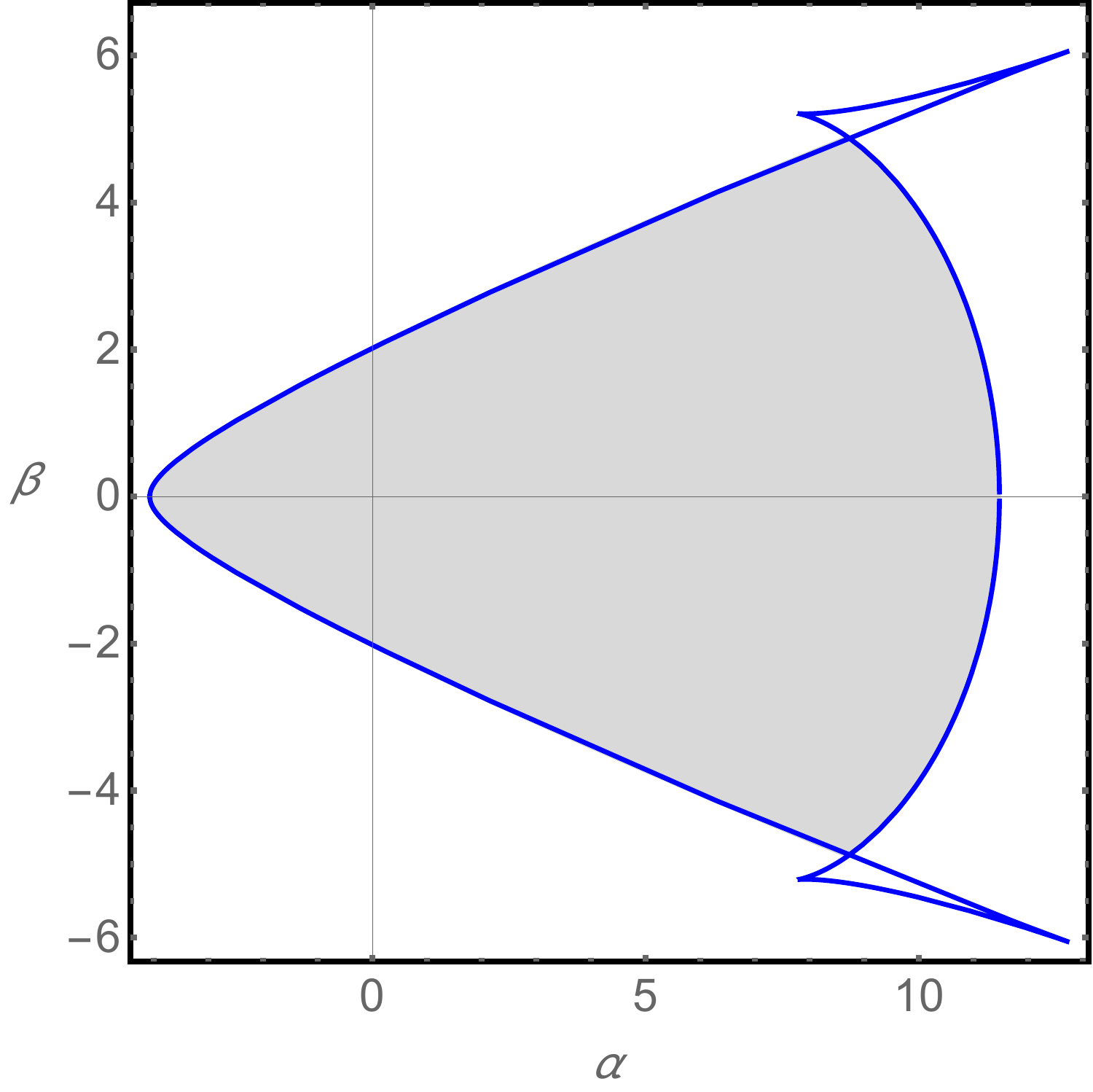}}
    \caption{Shadow cast by running-$G$ Kerr black hole. For $a_*<a_c\approx 2.343$, the shadow is the standard quasi-circular contour, as shown in (a).
     The critical behavior with $a_*=a_c$ is shown in (b), and for $a_*>a_c$, there are cuspy behaviors for the shadow as shown in (c) and (d).}
    \label{fig:RGKerr}
\end{figure}

In the spacetime of a Kerr black hole with a running Newton coupling, the shadow is governed by the behavior of null geodesics. Due to the stationarity and axisymmetry of the background, photon trajectories possess two conserved quantities: the energy $E = -p_t$ and the axial angular momentum $L = p_\phi$. Employing the Hamilton-Jacobi equation, the radial and angular equations of motion can be decoupled as follows:
\begin{equation}
\Sigma \dot{r} = \pm \sqrt{\mathcal{R}(r)}, \quad \Sigma \dot{\theta} = \pm \sqrt{\Theta(\theta)},
\end{equation}
with
\begin{equation}
\begin{aligned}
\mathcal{R}(r) &= \left[(r^{2} + a^{2})E - a L\right]^{2} - \Delta \left[ \mathcal{K} + (L - a E)^{2} \right], \\
\Theta(\theta) &= \mathcal{K} + \left( a^{2} E^{2} - \frac{L^{2}}{\sin^{2}\theta} \right) \cos^{2}\theta,
\end{aligned}\label{R-theta}
\end{equation}
where $\mathcal{K}$ is the Carter separation constant. The dots in $\dot{r}$ and $\dot{\theta}$ represent taking the derivative with respect to the affine parameter.
Dimensionless constants of motion can be defined as
\begin{equation}
\xi \equiv \frac{L}{E}, \quad \eta \equiv \frac{\mathcal{K}}{E^{2}}.
\end{equation}
The black hole shadow can be determined by the unstable circular photon orbits, that satisfy
\begin{equation}
\R(r) = 0, \quad \R'(r) = 0, \quad \R''(r) > 0.
\end{equation}
Here we use $r$ to represent the radius of the unstable circular orbits.
Solving the first two equations yields 
\begin{eqnarray}
\xi &=& \frac{1}{a} \left( r^2 + a^2 - \frac{4r\Delta(r)}{\Delta'(r)} \right),\\
\eta &=& 	\frac{16r^2\Delta(r)}{[\Delta'(r)]^2} - \frac{1}{a^2} \left( r^2  - \frac{4r\Delta(r)}{\Delta'(r)} \right)^2.
\end{eqnarray}
Assuming the observer is at infinity with inclination angle $\theta_0$, we can rewrite the black hole shadow in the celestial coordinates $(\alpha, \beta)$.
In the observer's local sky, the coordinates $(\alpha, \beta)$ are given by
\begin{eqnarray}
\begin{split}
	\alpha &= -\frac{\xi}{\sin\theta_0},\\
	\beta &= \pm \sqrt{\eta + a^2\cos^2\theta_0 - \xi^2\cot^2\theta_0}.
\end{split}\label{eq:ab0}
\end{eqnarray}
For an observer at the equatorial plane ($\theta_0=\pi/2$), the celestial coordinate $(\alpha, \beta)$ can be explicitly expressed as \cite{Chen:2025jay}
\begin{eqnarray}
\begin{split}
	\alpha &= \frac{3G(z)Mr^2 - r^3 -a^2(G(z)M + r)}{a(G(z)M - r) },\\
	\beta &=\pm \frac{\sqrt{4a^2G(z)Mr^3  - r(3G(z)M - r)^2}}{a(r - G(z)M)}.
\end{split}\label{eq:ab1}
\end{eqnarray}
Figure.~\ref{fig:RGKerr} illustrates the shadow of a Kerr black hole with a running gravitational constant $G(z)$. The panels (a)-(d) depict how the shadow's shape evolves as the normalized spin parameter $a_*$ increases. For the model we are considering, there is cuspy behavior for the black hole shadow for a specific range of parameter $a_*$.
In panel (\ref{fig:RGKerr_1}), where $a_* < a_c$ (with $a_c \approx 2.343$), the shadow exhibits the characteristic quasi-circular silhouette typical of a standard Kerr black hole. 
As the spin approaches the critical value $a_c$, the shadow undergoes a distinct morphological transition, shown in panel (\ref{fig:RGKerr_2}). For supercritical spins $a_* > a_c$, displayed in panels (\ref{fig:RGKerr_3}) and (\ref{fig:RGKerr_4}), the unstable circular orbits develop self-intersecting behavior, and correspondingly the shadow develops pronounced cusps, which possess sharp and pointed features that deviate markedly from the smooth quasi-circular contour.

\subsection{Topological charge and topological phase transition}

As discussed in Section~\ref{KZ}, the global structure of the unstable circular orbits can be characterized by a topological charge derived via the Gauss-Bonnet theorem. Because this topological charge for the shadow boundary is defined independently of the underlying spacetime geometry, the concept can be readily generalized to other compact objects.

In the context of the running-$G$ Kerr black hole, we can define the topological charge following the same logic and study its properties.
Using the Gauss-Bonnet theorem, the topological charge $\delta$ associated with the shadow boundary can be defined as
\begin{equation}
\delta = \frac{1}{2\pi} \left( \oint  \kappa\dd l + \sum_i \Delta \theta_i \right).\label{delta2}
\end{equation}
where $\kappa$ is the local curvature along the smooth segments of the boundary, and $\Delta \theta_i$ are the exterior angles at any non-differentiable points. $\kappa$ is the inverse of the curvature radius $1/R$.
For the shadow boundary parameterized as $(\alpha(r), \beta(r))$, the curvature is given by
\begin{equation}
\kappa =\frac{\alpha' \beta'' - \beta' \alpha''}{\left(\alpha'^2 + \beta'^2\right)^{3/2}},
\end{equation}
where the prime denotes differentiation with respect to the unstable circular orbits $r$.
The line element can be written as
\begin{equation}
\dd l=\sqrt{\alpha'^2 + \beta'^2}\, \dd r.
\end{equation}
Therefore, we have
\begin{equation}
\oint \kappa \, \dd l
=\oint
\frac{\alpha' \beta'' - \beta' \alpha''}
{\alpha'^2 + \beta'^2}
\, \dd r.
\end{equation}
Hence, the topological charge defined via the Gauss-Bonnet theorem becomes
\begin{equation}
\delta=\frac{1}{2\pi}\left(
\oint
\frac{\alpha' \beta'' - \beta' \alpha''}
{\alpha'^2 + \beta'^2}
\, \dd r + \sum_i \Delta \theta_i \right)
\end{equation}
For a smooth, simple closed curve as shown in Fig. (\ref{fig:RGKerr_1}), we can use the slope of the shadow boundary
\begin{equation}
\mathcal{F} = \frac{\dd\beta}{\dd\alpha}\label{eq:F}
\end{equation}
to evaluate $\delta$. Differentiating $\mathcal{F}$, we have
\begin{equation}
	\dd(\arctan \mathcal{F})=\frac{1}{1+\mathcal{F}^2}\, \dd\mathcal{F}.
\end{equation}
From definition \eqref{eq:F}, we have
\begin{equation}
\dd \mathcal{F}=\frac{\alpha'\beta'' - \beta'\alpha''}{\alpha'^2}\, \dd r,
\end{equation}
and
\begin{equation}
1+\mathcal{F}^2=\frac{\alpha'^2+\beta'^2}{\alpha'^2}.
\end{equation}
Therefore, we obtain
\begin{equation}
\dd(\arctan \mathcal{F})
=
\frac{\alpha'\beta'' - \beta'\alpha''}
{\alpha'^2+\beta'^2}
\, \dd r, 
\end{equation}
meaning that we can rewrite the topological charge as the change of the argument along the shadow boundary.

For a smooth shadow (e.g. $a_* < a_c$ in the running-$G$ case shown in Fig. (\ref{fig:RGKerr_1}), the boundary is differentiable everywhere and the second term in \eqref{delta2} is absent. 
Owing to the $\mathcal{Z}_2$  symmetry of the shadow, the expression acquires an additional factor of 2, simplifying the topological charge to 
\begin{equation}
	\delta = \frac{1}{\pi}(\arctan\mathcal{F}_l-\arctan\mathcal{F}_r)\,,\label{delta3}
\end{equation}
where $\mathcal{F}_l$ and $\mathcal{F}_r$ denote the slopes at the leftmost and rightmost points of the shadow, respectively.
At these extreme points, the slope becomes infinite, which means that we have $\mathcal{F}_l = -\infty$ and $\mathcal{F}_r = +\infty$. Consequently, equation \eqref{delta3} along with
\begin{equation}
\arctan(+\infty) = \frac{\pi}{2}, \qquad \arctan(-\infty) = -\frac{\pi}{2},
\end{equation}
gives 
\begin{equation}
	\delta = \frac{1}{\pi}\left[\frac{\pi}{2} - \Big{(}-\frac{\pi}{2}\Big{)}\right] = 1.
\end{equation}
 Hence, due to the smoothness of the shadow boundary, the running-$G$ Kerr shadow with $a_* < a_c$ belongs to the topological class $\delta = +1$. This is the same for the normal Kerr shadow and the KZ black hole with $\e$ greater than the critical value of the deformation parameter $\e_c$.

\begin{figure}[H]
\centering
  \includegraphics[width=0.7\textwidth]{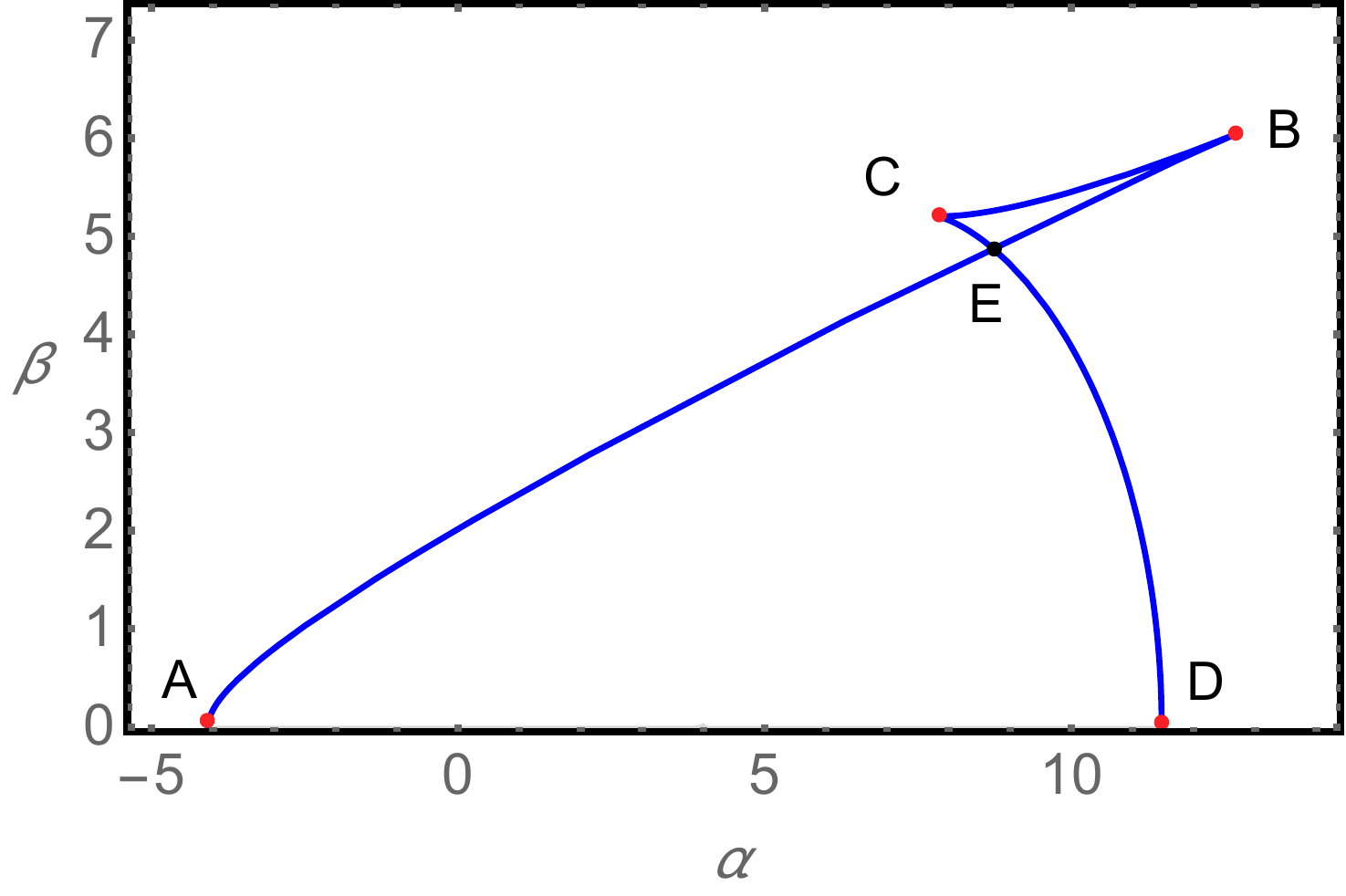}
  \caption{An illustration of the caluation of topological charge. The smooth part contains three segments: $A-B$, $B-C$ and $C-D$. Moreover, there are two exterior angles at points $B$ and $C$, each equals $\Delta \theta=-\pi$. }\label{fig:top}
\end{figure}

As $a^*$ increases beyond the critical threshold $a_c \approx 2.343$, the shadow undergoes a profound morphological transformation. Specifically, as depicted in Fig.~(\ref{fig:RGKerr_3}) and~(\ref{fig:RGKerr_4}), swallowtail features manifest along the unstable circular orbits. The vertices of these structures, marked as singular points $B$ and $C$ in Fig.~\ref{fig:top}, correspond to discontinuities in the tangent vector. These discontinuities introduce a discrete jump in the exterior angle, thereby necessitating the inclusion of the second term in Eq.~\eqref{delta2}.
Let us first look at the smooth segments that can be treated by the integration.
Exploiting the $\mathcal{Z}_2$ symmetry, the smooth part can be written as the summation of the three segments shown in Fig. \ref{fig:top}, i.e.,
\begin{equation}
\begin{split}
&\frac{1}{2\pi}\left(
\oint
\frac{\alpha' \beta'' - \beta' \alpha''}
{\alpha'^2 + \beta'^2}
\, \dd r \right)\\
&= \frac{1}{\pi}\left[(\arctan\mathcal{F}_D-\arctan\mathcal{F}_C)+(\arctan\mathcal{F}_C-\arctan\mathcal{F}_B)+(\arctan\mathcal{F}_B-\arctan\mathcal{F}_A) \right]\\
	&=\frac{1}{\pi}(\arctan\mathcal{F}_D-\arctan\mathcal{F}_A)=+1\,.
\end{split}\nonumber
\end{equation}
Note that the two segments approach points $B$ and $C$ with the same slope, which is the key reason why we are only left with the contributions from points $A$ and $D$.
Each exterior angle at points $B$ and $C$ contributes an angle jump of $\Delta \theta = -\pi$.
So for each shadow shown in Fig. (\ref{fig:RGKerr_3}) and (\ref{fig:RGKerr_4}), incorporating these contributions and the $\mathcal{Z}_2$ symmetry, the topological charge becomes
\begin{equation}
\delta = \frac{1}{\pi} \bigl( \pi + \Delta\theta_B + \Delta\theta_C \bigr) = \frac{1}{\pi} (\pi - \pi - \pi) = -1.
\end{equation}
Thus, the topological charge flips from $+1$ to $-1$ as the shadow transitions from a smooth quasi-circular contour to a cuspy one. 

This is not a continuous deformation but a genuine topological phase transition, where the self-intersecting photon orbit alters the shadow's genus. The change in $\delta$ reflects a fundamental reorganization of the shadow's global structure and places cuspy shadows in a distinct topological class from their smooth counterparts.

\subsection{Equal-area law}

The self-intersecting unstable circular orbits bear a striking resemblance to the swallowtail behavior observed in the free energy of thermodynamic systems.
Just like Maxwell's equal-area law, there are similar equal-area law for the unstable circular orbits with swallowtail behavior.
Analogously, the equal-area law governing the morphology of cuspy black hole shadows also offers a rigorous method to determine the self-intersection point where two distinct circular photon orbits project to the same celestial coordinates.

The equal-area law is a general characterization of systems with self-intersections.
From Fig. \ref{fig:top}, the curve develops a closed contour $\mathcal{C}$: $E-B-C-E$; thus, we have the loop integral of $\dd \beta$ vanishes along the closed contour
\begin{equation}
	\oint_{\mathcal{C}} \dd \beta=0\,.
\end{equation}
Using the definition of the slope $\mathcal{F}$, we have
\begin{equation}
\oint_{\mathcal{C}} \mathcal{F}\, \dd\alpha = 0.	\label{eq:oint}
\end{equation}

\begin{figure}[H]
\begin{center}
  \subfloat[$a>a_c$\label{fig:RG-area1}]{\includegraphics[width=0.45\textwidth]{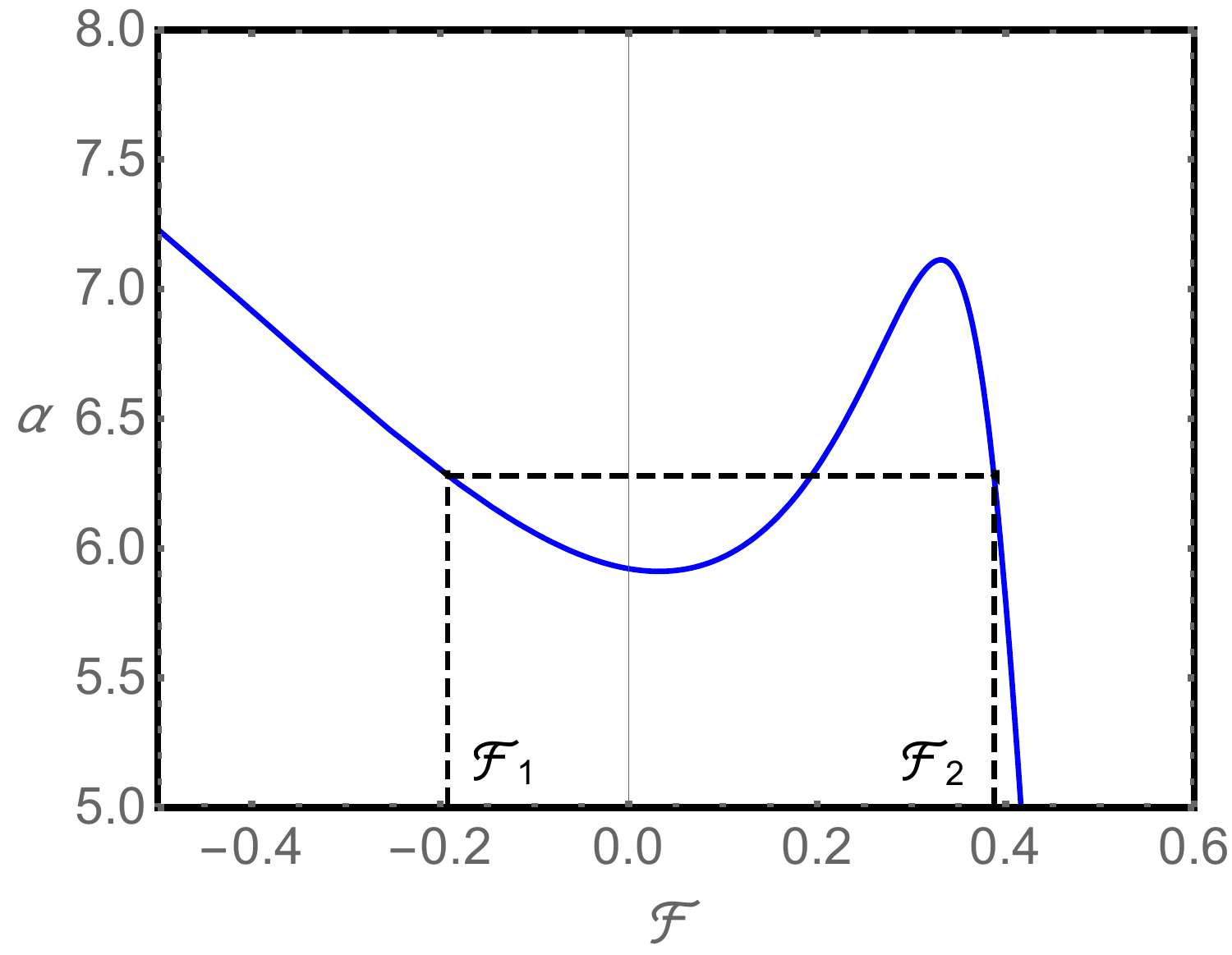}}~~~~
  \subfloat[$a=a_c$\label{fig:RG-area2}]{\includegraphics[width=0.45\textwidth]{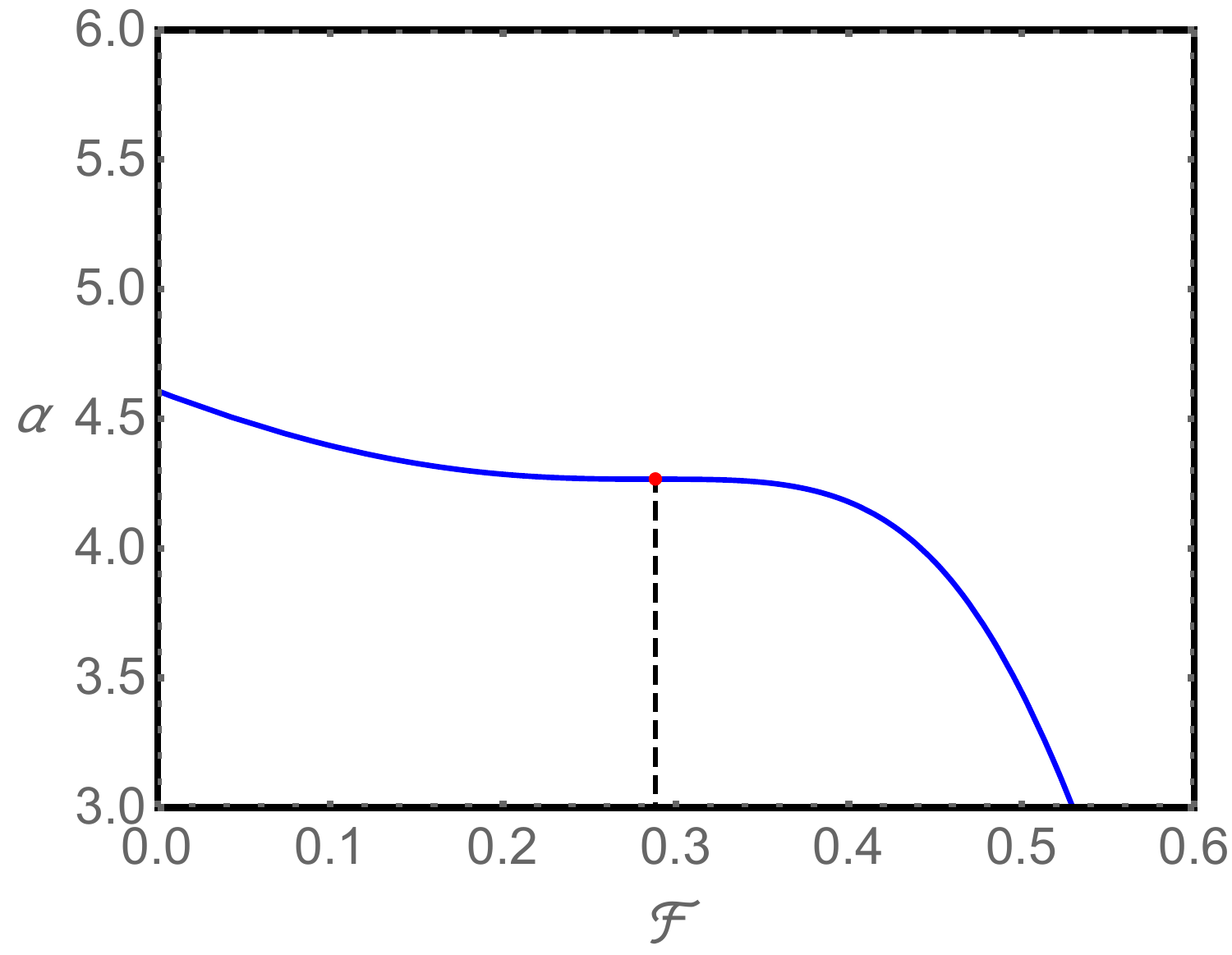}}
  \caption{The equal-area law for the cuspy shadow of the running-$G$ Kerr black hole. 
  Above criticality $a>a_c$, there is an equal-area law in the $\alpha-\mathcal{F}$ diagram.
  At the critical point, the interval $[\mathcal{F}_1,\mathcal{F}_2]$ collapses to zero, and the discontinuous jump reduces to an inflection point.
  }\label{fig:RG-area}
\end{center}
\end{figure}

At the self-intersection point $E=(\alpha_i, \beta_i)$, two distinct photon orbits (with radii $r_1$ and $r_2$) coincide on the celestial sphere. 
Consequently, the curve $\alpha(\mathcal{F})$ becomes multi-valued in the interval between $\mathcal{F}_1 = \mathcal{F}(r_1)$ and $\mathcal{F}_2 = \mathcal{F}(r_2)$. 
Applying condition \eqref{eq:oint} to a contour that traverses the upper and lower branches of the multi-valued region yields the equal-area law
\begin{equation}
\int_{\mathcal{F}_1}^{\mathcal{F}_2} \alpha\,\dd\mathcal{F} = \alpha_i \cdot (\mathcal{F}_2 - \mathcal{F}_1)\,.\label{eq:area}
\end{equation}
From a geometric perspective, Eq.~\eqref{eq:area} dictates that the two signed areas bounded by the non-monotonic curve $\alpha(\mathcal{F})$ and the vertical line $\alpha = \alpha_i$ are identical, giving rise to the well-known ``equal-area law". As depicted in Fig.~\ref{fig:RG-area}, this principle is numerically verified for the running-$G$ Kerr black hole. Crucially, this relationship emerges as a direct consequence of the closed nature of the parameterized shadow contour, remaining entirely independent of the specific spacetime metric or any underlying thermodynamic analogies.

As the spin parameter decreases toward the critical value $a_c$, the two areas in the equal-area construction shrink monotonically and vanish exactly at $a_c = 2.343$, where the curve $\alpha(\mathcal{F})$ develops an inflection point. 
The $\alpha-\mathcal{F}$ relation shown in Fig. (\ref{fig:RG-area2}) exactly mirrors the situation illustrates in Fig. (\ref{fig:RGKerr_2}).
For subcritical spins ($a_* < a_c$), the curve becomes monotonic, and the equal-area construction no longer applies, consistent with the absence of self-intersection.  
The equal-area law on the black hole shadow boundary thus provides a robust and predictive tool for analyzing shadow morphology in non-Kerr spacetimes, establishing a universal principle governing the black hole optics.

\subsection{Critical phenomena}

The transition from a smooth to a cuspy shadow is not merely a morphological change but constitutes a genuine critical phenomenon, characterized by universal scaling behavior and critical exponents. The critical point, the exact threshold where cusps first appear, occurs when the two equal areas in the area law construction vanish simultaneously, as shown in Fig. (\ref{fig:RG-area2}). 
%At the critical point, two distinct branches of unstable circular photon orbits merge into a single degenerate orbit.
Geometrically, this corresponds to an inflection point in the $(\alpha, \mathcal{F})$ plane, satisfying the conditions
\begin{equation}
\left( \frac{\partial \alpha}{\partial \mathcal{F}} \right)_{a=a_c} = 0, \qquad \left( \frac{\partial^2 \alpha}{\partial \mathcal{F}^2} \right)_{a=a_c} = 0. \label{eq:critical}
\end{equation}
For the running-$G$ Kerr black hole with fixed running coupling parameters, the critical spin is determined to be $a_c \approx 2.343$, with a corresponding critical celestial coordinate of $\alpha_c/(G_0M) \approx 2.088$. Below the critical spin ($a_* < a_c$), the shadow is smooth; above it ($a_* > a_c$), the shadow develops cusps, and the self-intersection becomes progressively more pronounced as $a_*$ increases.

To quantitatively characterize the critical behavior, we introduce an order parameter, as the separation between the two slope values at the self-intersection point
\begin{equation}
	\Delta \mathcal{F} = \mathcal{F}_2 - \mathcal{F}_1.
\end{equation}
$\Delta \mathcal{F}$ vanishes at the critical point and grows as the system moves deeper into the cuspy phase, which is illustrated in Fig. \ref{fig:RG-area}.
Near the critical point, we have
\begin{equation}
\Delta \mathcal{F} \to 0 \quad  \text{as}  \quad   a \to a_c.
\end{equation}
The order parameter exhibits power-law scaling with the distance from criticality
\begin{equation}
\Delta \mathcal{F} \sim \left( a_* - a_c \right)^{\zeta_1}, \qquad \Delta \mathcal{F} \sim \left( \frac{\alpha - \alpha_c}{G_0 M} \right)^{\zeta_2},
\end{equation}
where $\zeta_1$ and $\zeta_2$ are the critical exponents governing the scaling with respect to the spin parameter and the celestial coordinate, respectively.

\begin{figure}[H]
\begin{center}
  \subfloat[Critical exponent $\zeta_1$]{\includegraphics[width=0.45\textwidth]{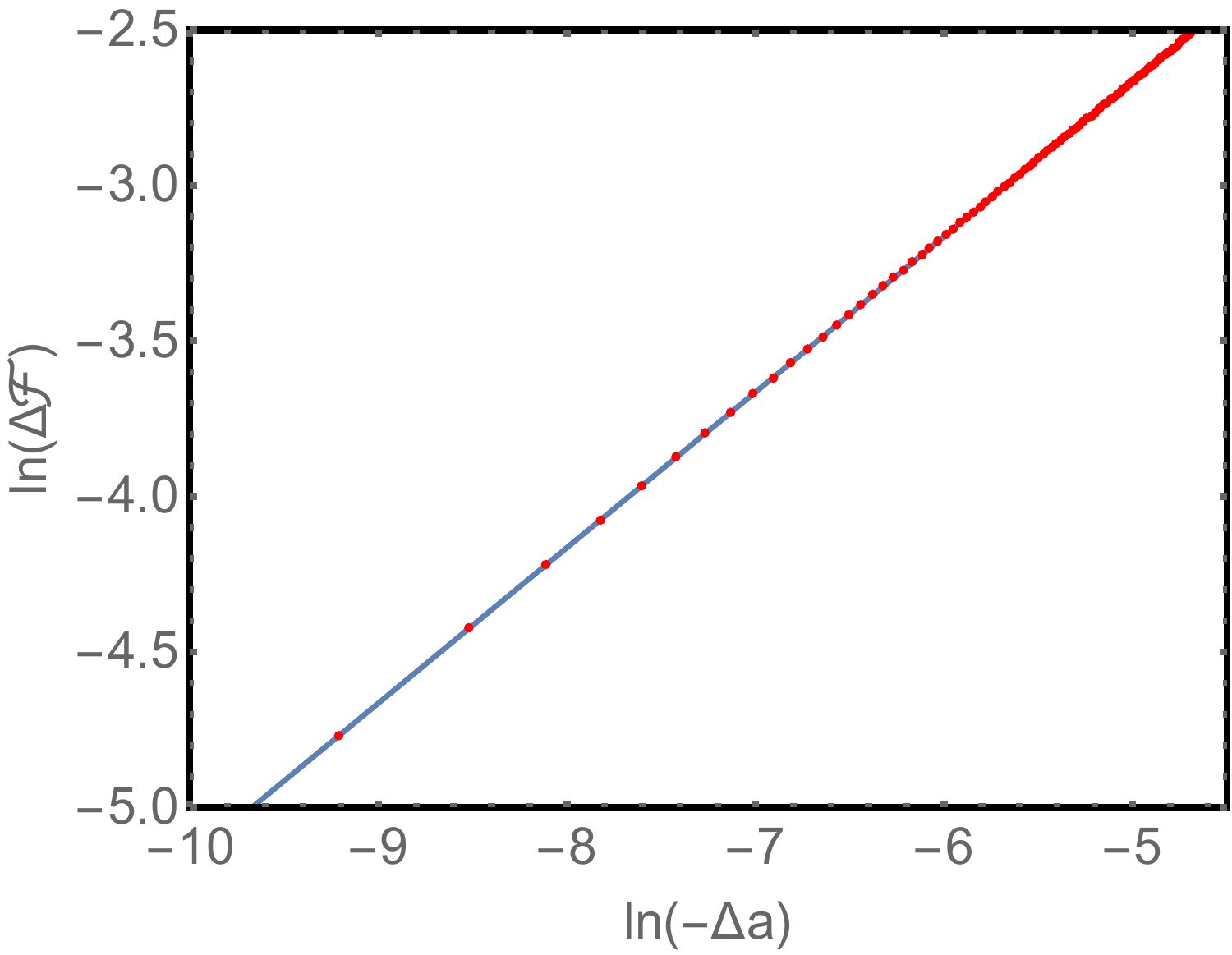}}\quad
  \subfloat[Critical exponent $\zeta_2$]{\includegraphics[width=0.45\textwidth]{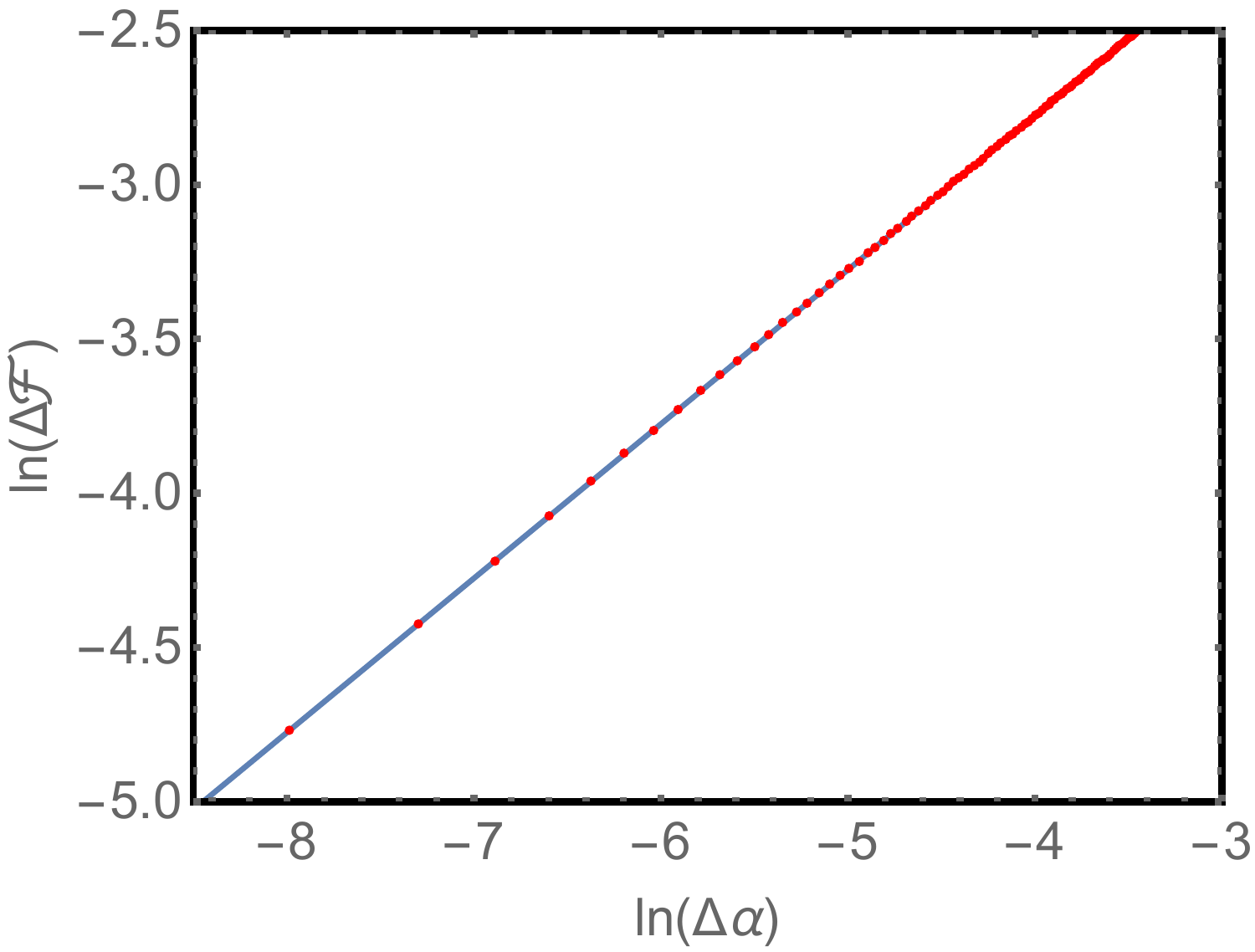}}
  \caption{Critical behavior of the running-$G$ Kerr black hole. 
  The critical exponents are $\zeta_1=\zeta_2=1/2$. 
  The numerical results near the criticality (represented by red dots) exhibit a scaling behavior consistent with the critical exponents, as validated by their close proximity to the reference blue lines of slope $1/2$.
  }\label{fig:KZ-ce}
\end{center}
\end{figure}

Our numerical analysis of the running-$G$ Kerr black hole reveals that the order parameter obeys these scaling laws with remarkable precision, as illustrated in Fig.~\ref{fig:KZ-ce}. Performing log-log fits of $\Delta \mathcal{F}$ versus $(a_* - a_c)$ and $(\alpha - \alpha_c)$ yields slopes of approximately $0.499$ and $0.499$, strongly indicating
\begin{equation}
\zeta_1 = \zeta_2 = \frac{1}{2}. 
\end{equation}
The exponent $1/2$ originates from the local degeneracy structure of the mapping near the critical point.

The critical exponent $\zeta = 1/2$ is of profound significance. It is the hallmark of the mean-field universality class, which encompasses a wide range of physical systems, including the Curie-Weiss model of ferromagnetism, the van der Waals liquid-gas transition, and Landau's theory of phase transitions. 
In all such systems, the order parameter scales as $(T_c - T)^{1/2}$ near the critical temperature. 
This universality implies that the critical exponent $1/2$ should be independent of the specific details of the black hole model. Whether one considers the KZ black hole or the running-$G$ Kerr black hole, the onset of cusp formation should always exhibit the same mean-field scaling behavior. 
This suggests that the critical phenomenon is deeply rooted in the geometry of photon orbits rather than in a specific black hole model.

\section{Universality of the cuspy shadow}
\label{universal}

As already implied in the previous section, he three features characterizing the cuspy shadow are model-independent, suggesting the existence of an underlying universal mechanism. 
While it is important to establish these phenomena for different models, the true power of these results lies in their generality. 
In this section, we further demonstrate that those phenomena are not coincidental features of specific black hole models, but rather manifestations of the universality principle governing cusp formation in compact object shadows.

The aim of this section is twofold. 
First, we identify the geometric and topological mechanisms that underlie this universality, showing that cusp formation is inherently tied to the emergence of the self-intersection structure in the space of unstable circular orbits.
Those three phenomena are dictated solely by the global behavior of the shadow rather than its detailed metric functions.
Second, we provide a nontrivial test of this universality by examining a completely different class of compact objects: rotating traversable wormholes. 
If the cuspy shadow phenomenon is truly universal, it should persist even in horizonless spacetimes. By demonstrating that rotating wormholes also exhibit the same topological transition, equal-area law, and critical exponent $1/2$, we establish that these features constitute a robust, model-independent fingerprint of strong-field gravity beyond the Kerr paradigm.

\subsection{The universal mechanism behind the cusp formation}

%Those three phenomena originate from the intrinsic geometric and topological properties of the unstable circular orbits when it develops a self-intersecting structure. 
Now, we analyse each phenomenon separately and demonstrate that its origin is completely independent of the detailed form of the metric, thus establishing the universality of cuspy shadows.

\subsubsection*{The topological charge flip}

\begin{figure}[H]
\centering
\begin{tikzpicture}[scale=1]
\begin{scope}[xshift=-3cm]
\draw[thick,<-] (0,0) circle (1);
\draw[thick,<-] (0.8,-0.2) arc (-15:15:0.8);
\node at (0,-2) {$\delta = +1$};
\node at (0,0) {\textcolor{red}{$+1$}};
%\node at (0,1.6) {Single loop};
\end{scope}
\begin{scope}[xshift=3cm]
\draw[thick,<-] (-1,0.45) circle (0.8);
\draw[thick,<-] (1,0.45) circle (0.8);
\draw[thick,<-] (0,-0.8) circle (0.8);
\draw[thick,<-] (0.6,-1) arc (-15:15:0.8);
\draw[thick,<-] (-1.4,0) arc (230:200:0.8);
\draw[thick,->] (1.4,0) arc (-50:-20:0.8);
\node at (-1,0.45) {\textcolor{blue}{$-1$}};
\node at (1,0.45) {\textcolor{blue}{$-1$}};
\node at (0,-0.8) {\textcolor{red}{$+1$}};
\node at (0,-2) {$\delta = -1$};
%\node at (0,2.0) {Three loops};
\end{scope}
\end{tikzpicture}
  \caption{An illustration of the topological charge. Stipulating the circle shown on the left panel has winding number $\delta=+1$, the curve on the right panel has winding number $\delta=-1$.  }\label{fig:circles}
\end{figure}
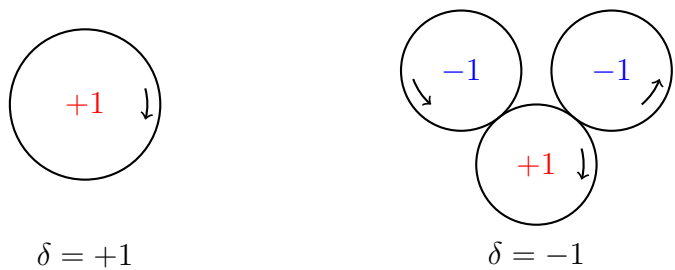

The universality of the topological charge transition from $+1$ to $-1$ upon the emergence of a swallowtail structure in the unstable circular orbits stems from the fact that this phenomenon is governed exclusively by the global geometric properties of the geometry, rather than its specific details. 
The topological charge defined via the Gauss-Bonnet theorem \eqref{delta2} is a purely geometric invariant that depends only on the total turning angle of the tangent vector along the closed curve and the discrete angle deficits at any singular points. 
For any smooth Jordan curve, the curvature integral equals $2\pi$, and there are no angle deficits, so we always have 
\begin{equation}
	\delta = +1\,.
\end{equation}
This holds for any shadow shown in the left panel of Fig. \ref{fig:circles}, irrespective of the specific metric functions.

However, when the swallowtail structure forms, it introduces two extra circles with discontinuous jumps in the tangent direction.
The two extra circles alter the global structure of the curve and are the key reason for the change in the topological charge.
As illustrated in the right panel of Fig. \ref{fig:circles}, the bear-head-shaped boundary carries a topological charge that differs from that of a simple Jordan curve.
It follows that, upon the formation of a pair of swallowtail structures, the topological charge defined by the Gauss-Bonnet theorem always transitions to $-1$. This result holds regardless of any discontinuous jumps or the detailed structure of the metric.

This argument is based only on the global structure of the shadow. 
%No reference is made to the underlying spacetime geometry or the specific mechanism that generates the cusps. 
Once the cusp/swallowtail structure is realized by the photon-orbit map, the resulting topological transition is controlled by the global structure of the shadow boundary rather than by the detailed metric functions.
Therefore, any compact object whose shadow boundary exhibits a self-intersecting swallowtail structure must undergo a topological phase transition from $\delta = +1$ to $\delta = -1$. This universality makes the topological charge a robust, model-independent characterization of the non-Kerr paradigm.

\subsubsection*{The equal-area law}

%The equal-area law also follows from a purely geometric condition that is insensitive to the metric details. 
The equal-area law is geometric in form, but in the present context, it acts as a constraint on the self-intersecting shadow boundary generated by photon-orbit dynamics. Its metric independence is precisely what makes it a universal relation for cuspy shadows.
Consider a shadow boundary that self-intersects at a point $(\alpha_i,\beta_i)$. 
The integral of $\dd\beta$ along the closed swallowtail must vanish.
Along this loop, we always have
\begin{equation}
\oint \dd\beta = 0 \quad\Longrightarrow\quad \oint \mathcal{F}\,\dd\alpha = 0,
\end{equation}
with $\mathcal{F} = \dd\beta/\dd\alpha$. 
Integration by parts yields the equal-area condition
\begin{equation}
\int_{\mathcal{F}_1}^{\mathcal{F}_2} \alpha \, \dd\mathcal{F} = \alpha_i\,(\mathcal{F}_2 - \mathcal{F}_1).
\end{equation}
It is worth noting that the slopes of the two branches differ at their intersection point $(\alpha_i,\beta_i)$. We denote these two slopes by $\mathcal{F}_1$ and $\mathcal{F}_2$, respectively.

%This derivation relies solely on the existence of a self-intersection within the physical shadow. 
This derivation relies on the existence of a self-intersection within the physical shadow boundary, which itself arises from the degeneracy of the photon-orbit projection map.
The law emerges as a direct consequence of the closed nature of the swallowtail structure. Consequently, any cuspy shadow, provided it corresponds to closed swallowtail configurations of unstable circular orbits, must satisfy the same equal-area law.

\subsubsection*{Universal critical exponent}

Near the critical point where the cusps first appear, the shadow boundary exhibits universal scaling behaviour. 
The critical point is characterised by the vanishing of the first and second derivatives of $\alpha$ with respect to $\mathcal{F}$:
\begin{equation}
\frac{\partial\alpha}{\partial\mathcal{F}}\Bigg{|}_{\xi=\xi_c}= 0,\qquad 
\frac{\partial^2\alpha}{\partial\mathcal{F}^2 }\Bigg{|}_{\xi=\xi_c}= 0.\label{eq:condition}
\end{equation}
where we use $\xi$ to denote the control parameter, like the spin, deformation, or other parameters.
Moreover, $\xi_c$ is used to denote the critical value of the control parameter.
These conditions imply that the local expansion of $\alpha(\mathcal{F},\xi)$ around the critical point can be written as
\begin{equation}
\alpha - \alpha_c \approx A\cdot(\mathcal{F} - \mathcal{F}_c)^3 + B\cdot(\xi-\xi_c)+ C\cdot(\xi-\xi_c)(\mathcal{F} - \mathcal{F}_c)+\cdots\,,\label{eq:expand}
\end{equation}
with constants $A$, $B$ and $C$.
The order parameter $\Delta\mathcal{F}$ defined before is proportional to $\mathcal{F} - \mathcal{F}_c$ near the critical point. 
Therefore, the first equation in \eqref{eq:condition} reduces to a quadratic equation in $\Delta\mathcal{F}$. The solution is  the square root of the distance from criticality
\begin{equation}
	\Delta\mathcal{F} \sim |\xi - \xi_c|^{1/2}.
\end{equation}
Note that due to the presence of the term $B\cdot(\xi-\xi_c)$ in \eqref{eq:expand}, we also have 
\begin{equation}
\Delta\mathcal{F} \sim |\alpha - \alpha_c|^{1/2}.
\end{equation}
The exponent $1/2$ follows solely from the non-degenerate vanishing of the first two derivatives in Eq.~\eqref{eq:condition}; it is a direct consequence of the local geometry of the inflection point. No further details of the metric or the exact form of the shadow were entered.
It is worth noticing that this square-root scaling is the hallmark of the mean-field universality class, encountered in a wide variety of phase transitions. The critical exponents in the mean-field universality also originate from the same mechanism, and the universal exponents in shadows actually do not depend on the thermodynamic analogy.

%Its appearance here demonstrates that the bifurcation of unstable circular orbits belongs to the same universal class. The exponent $1/2$ confirmed numerically for the KZ black hole, the running-$G$ Kerr black hole providing strong evidence that the critical behaviour is truly universal.

The three phenomena discussed above, topological charge flip, equal-area law, and critical scaling with exponent $1/2$, form a coherent and universal description of cuspy shadow formation.
This universality does not depend on the specific feature of the metric, but is a generic consequence of the global structure of the null geodesics.
%This universality provides a powerful, model-independent probe of strong-field gravity.

\subsection{A non-trivial example: rotating traversable wormhole}

To ascertain whether the feature discussed above is widespread, we need to examine diverse classes, especially non-trivial cases beyond black holes. 
Let us test whether the universal characteristics apply to cuspy shadows cast by rotating traversable wormholes.
If it applies not only to black holes but also to compact objects, the arguments in this section would be more robust.

The metric of a rotating traversable wormhole is given by \cite{Teo:1998dp}
\begin{equation}
ds^{2} = -N^{2}\dd t^{2} + \left(1 - \frac{b}{r}\right)^{-1}\dd r^{2} + r^{2}K^{2}\left[\dd \theta^{2} + \sin^{2}\theta (\dd \phi - \omega \dd t)^{2}\right],
\end{equation}
where the metric functions are chosen to be
\begin{equation}
	N = \exp\left(-\frac{r_0}{r} - \lambda\frac{r_0^2}{r^2}\right),\quad b = r_0,\quad K = 1,\quad \omega = \frac{2J}{r^{3}}.
\end{equation}
with the redshift parameter $\lambda$. It was shown in \cite{Cheng:2026wyk} that the formation of cusps is controlled by the parameter $\lambda$.
The wormhole throat is located at $r = r_0$, with $b(r_0) = r_0$.
The shadow boundary is determined by two families of unstable circular photon orbits:

\begin{itemize}
  \item \noindent\textbf{Critical circular orbits outside the throat}. The unstable circular photon orbits correspond to $r > r_0$ satisfying $\R(r)=\R'(r) = 0$, which yields a parametric curve
\begin{equation}
	\begin{split}
		\alpha(r_{\text{ph}}) &= -\frac{1}{\sin\theta_{\text{ob}}} \cdot \frac{r^3 (r_0 r - r^2 + 2\lambda r_0^2)}{2J (r_0 r + 2 r^2 + 2\lambda r_0^2)}\bigg|_{r=r_{\text{ph}}}, \\[6pt]
\beta(r_{\text{ph}}) &= \pm \frac{r^3\sqrt{ 36 J^2 \sin^2\theta_{\text{ob}} \, N^{-2}(r) - (r_0 r - r^2 + 2\lambda r_0^2)^2 }}{2J (r_0 r + 2 r^2 + 2\lambda r_0^2) \sin\theta_{\text{ob}}} \bigg|_{r=r_{\text{ph}}}.
	\end{split}\label{eq:ab}
\end{equation}
$r_{\text{ph}}\in [r_0,\infty)$ is the running parameter characterizing unstable circular orbits radius.

\item \noindent\textbf{Critical orbits at the throat}.
At the throat $r = r_0$, the effective potential exhibits an extremum. 
The critical conditions yield an implicit relation between the impact parameters
\begin{equation}
(N_0^{2} - \omega_0^{2} r_0^{2} K_0^{2} \sin^{2} \theta_{\text{ob}}) \alpha^{2} - 2\omega_0 r_0^{2} K_0^{2} \sin \theta_{\text{ob}} \, \alpha - r_0^{2} K_0^{2} + N_0^{2} \beta^{2} = 0.\label{eq:throat}
\end{equation}
At the throat, we have
\begin{equation}
	N_0 = \exp\left[-\frac{r_0}{r_0} - \lambda\left(\frac{r_0}{r_0}\right)^2\right] = e^{-1-\lambda}, \quad K_0 = 1, \quad \omega_0 = \frac{2J}{r_0^3}.
\end{equation}
Thus, we can rewrite the wormhole throat orbits \eqref{eq:throat} as
\begin{equation}
	\beta = \pm N_0^{-1}\sqrt{ r_0^2\left(1+\omega_0 \sin\theta_{\text{ob}}\alpha\right)^2-\alpha^2 N_0^2}.\label{eq:throat0}
\end{equation}
\end{itemize}
%The shadow boundary is the superposition of the $(\alpha, \beta)$ curves from the two families of orbits \eqref{eq:ab} and \eqref{eq:throat0}.

\begin{figure}[H]
    \centering
    \subfloat[$\lambda=0.2$\label{fig:WH_1}]{\includegraphics[width=0.28\textwidth]{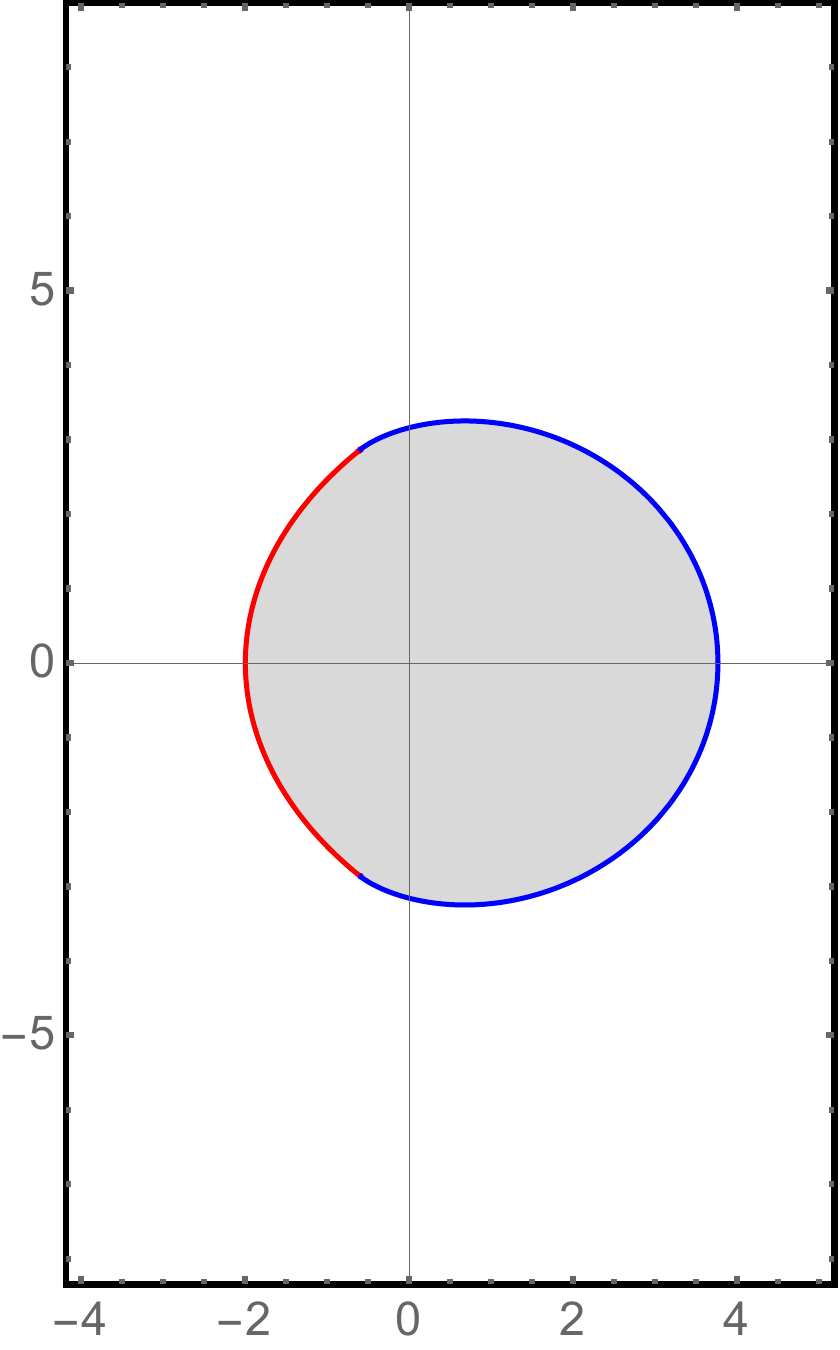}}
    \subfloat[$\lambda=\lambda_c=\frac{1}{4}(\sqrt{5}-1)$\label{fig:WH_2}]{\includegraphics[width=0.28\textwidth]{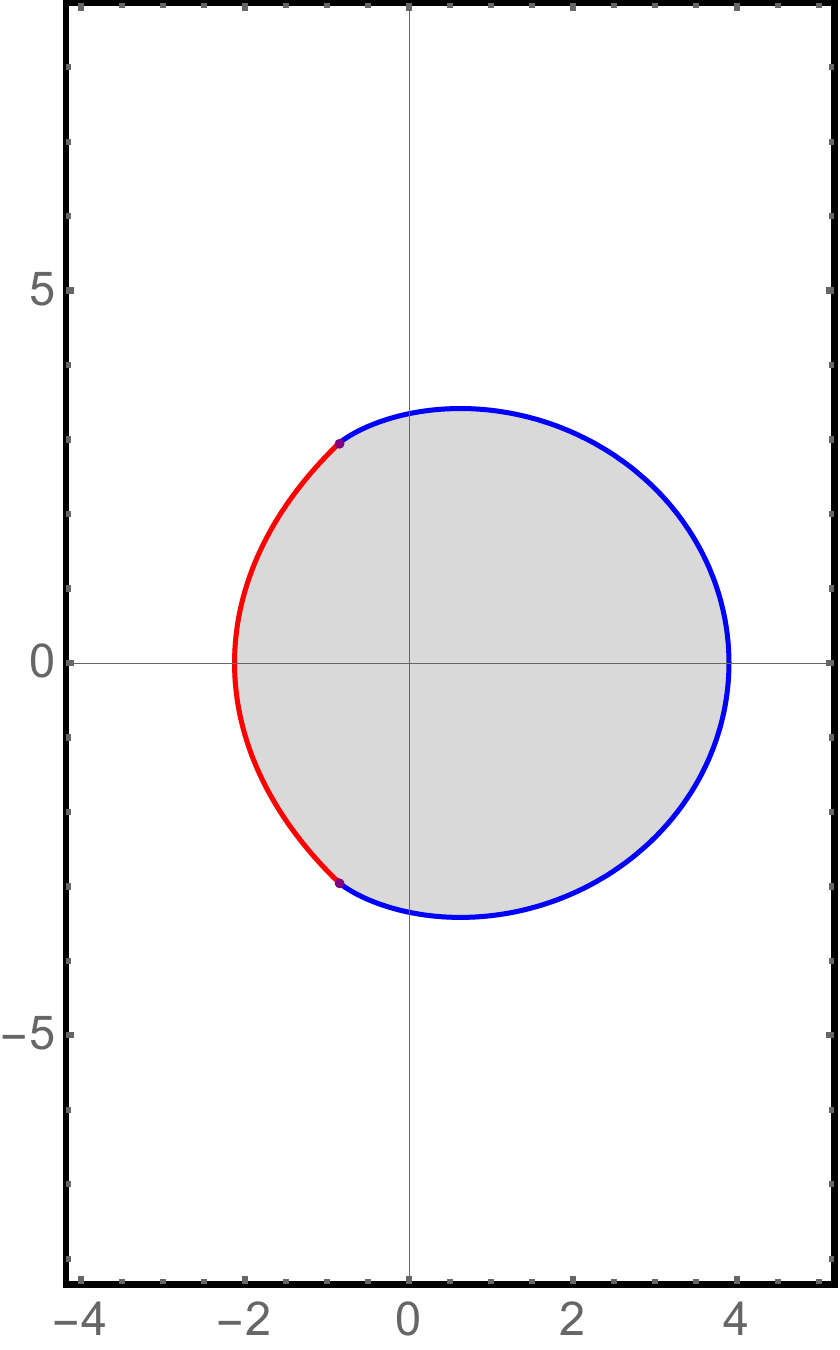}}
    \subfloat[$\lambda=1.2$\label{fig:WH_3}]{\includegraphics[width=0.28\textwidth]{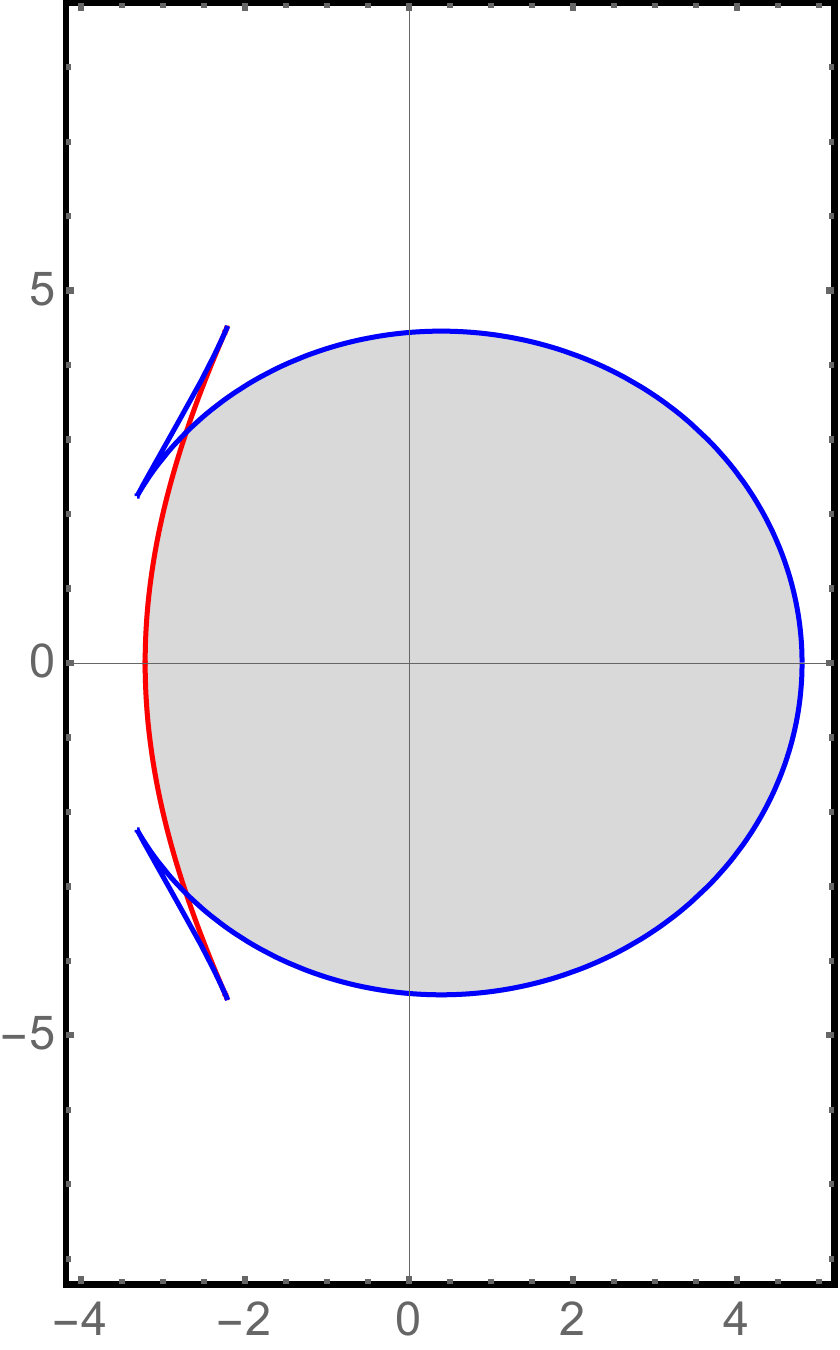}}
    \subfloat{\begin{tikzpicture}
    \draw [thick,->] (0,-1)--(0,1) node[above=5pt]{$\beta$};
    \draw [thick,->] (0,0)--(1,0) node[midway,below=3pt]{$\alpha$};
    \node at (0,-3.65) {};
    \end{tikzpicture}}
    \caption{Shadows for the rotating traversable wormhole with $a = 0.1$. 
    The blue and red curves approach each other with the same slope.
    For the redshift parameter $\lambda<\lambda_c$, the shadow is smooth.
    While the shadow develops a swallowtail structure for $\lambda>\lambda_c$. The critical value for $\lambda$ is $\lambda_c=\frac{1}{4}(\sqrt{5}-1)$, which labels the onset of the cusp formation.
    }
    \label{fig:WH}
\end{figure}
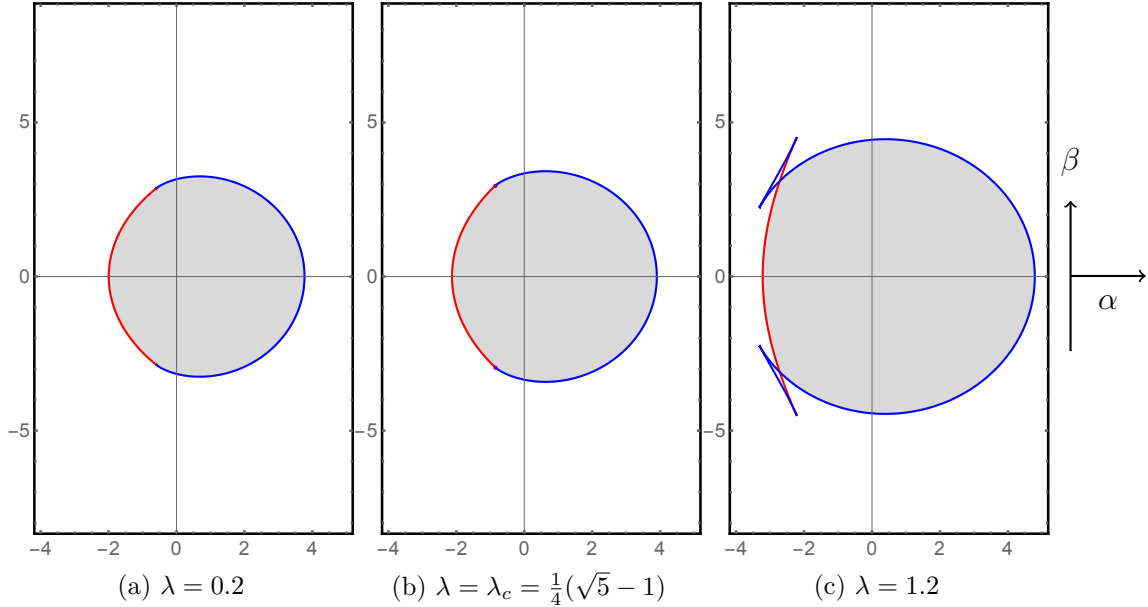

To facilitate numerical visualization, we scale all physical quantities by $GM$ to render them dimensionless, yielding $r_0 = 1$. The shadows cast by the rotating traversable wormhole with a spin parameter $a = 0.1$ are depicted in Fig.~\ref{fig:WH}. As illustrated, the shadow morphology depends critically on the redshift parameter $\lambda$.
It can be verified that the blue and red curves approach each other with the same slope.
Moreover, the curve is smooth in panel (\ref{fig:WH_1}), while there is an extra turning angle $\Delta\theta=-\pi$ in panel (\ref{fig:WH_3}).
So, for $\lambda < \lambda_c$, the shadow boundary is smooth. When $\lambda$ is larger than the critical value $\lambda_c = (\sqrt{5}-1)/4$, the shadow develops a self-interacting structure as shown in (\ref{fig:WH_3}). The critical point in panel (\ref{fig:WH_2}) marks the onset of cusp formation.

\subsubsection*{Properties of the cuspy wormhole shadow}

In the context of the shadow cast by a rotating traversable wormhole, we can define the topological charge of the combination of the unstable circular orbits and wormhole throat, as illustrated in Fig.~\ref{fig:WH}. 
According to the Gauss-Bonnet theorem, the topological charge $\delta$ is determined by the global properties of the curve. For a smooth, non-self-intersecting shadow boundary, as shown in panel~(\ref{fig:WH_1}), the topological charge is $\delta = +1$, placing it in the same topological class as the Kerr black hole shadow. 
However, when the redshift parameter $\lambda$ becomes larger than the critical value $\lambda_c$, the shadow boundary develops a self-intersecting structure and forms cusps, as illustrated in panel~(\ref{fig:WH_3}). 
In this case, two non-differentiable points emerge, each contributing an exterior angle of $\Delta \theta = -\pi$. Accounting for the $\mathcal{Z}_2$ symmetry, the total topological charge evaluates to $\delta = -1$. This reversal in topological charge is fully consistent with the behavior observed in the KZ black hole and the running-$G$ Kerr black hole, indicating that the wormhole shadow also undergoes a topological phase transition upon cusp formation. Crucially, the manifestation of this topological phase transition in compact objects beyond standard black hole models firmly confirms its universality.
And we can say that the topological charge is determined solely by the global structure of the shadow boundary and is independent of the specific metric form.

\begin{figure}[H]
\begin{center}
  \subfloat[$a=0.2$ and $\lambda=1$\label{fig:WH-area}]{\includegraphics[width=0.45\textwidth]{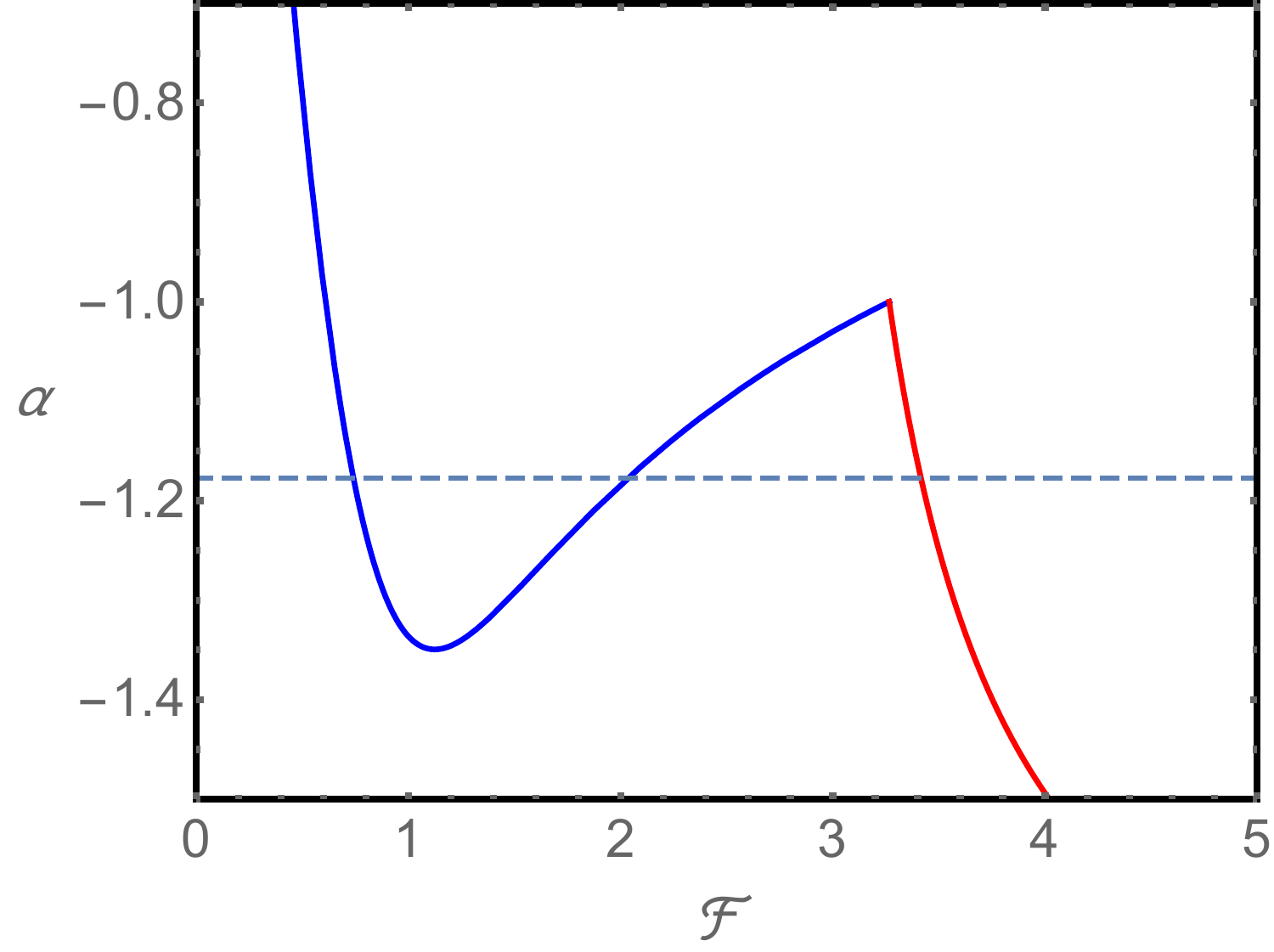}}~~~~
  \subfloat[Critical exponent $\zeta=1/2$ for $a=0.1$.\label{fig:WH-ce}]{\includegraphics[width=0.45\textwidth]{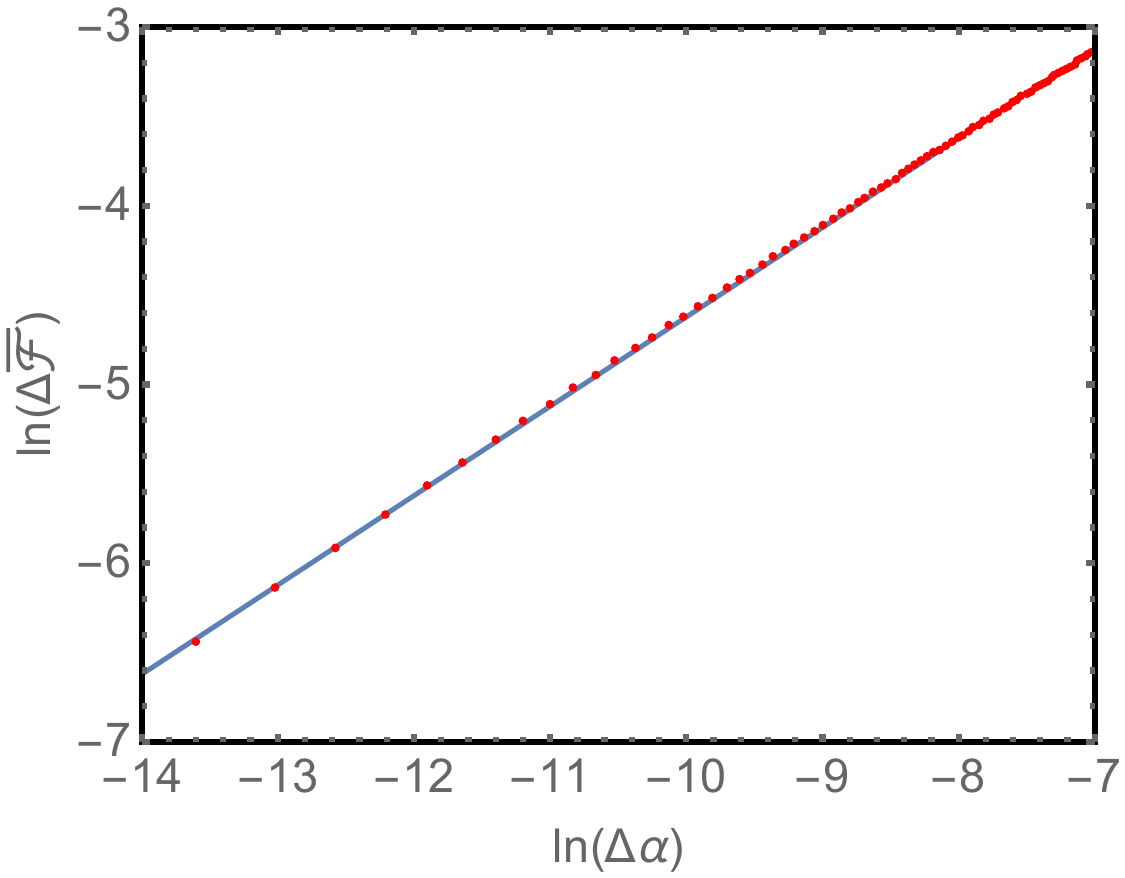}}
  \caption{Properties of the cuspy wormhole shadow. The figure in the left panel demonstrates the equal-area law, and the one in the right panel shows that the critical exponent is $1/2$.}\label{fig:WH-crit}
\end{center}
\end{figure}

When the wormhole shadow boundary self-intersects, as shown in Fig.~(\ref{fig:WH_3}), the boundary forms a closed loop. The integral of $\mathrm{d}\beta$ along this loop vanishes, and combined with the definition of the slope $\mathcal{F} = \mathrm{d}\beta/\mathrm{d}\alpha$, we obtain $\oint \mathcal{F} \,\mathrm{d}\alpha = 0$. Through integration by parts, this leads to the equal-area law:
\begin{equation}
\int_{\mathcal{F}_1}^{\mathcal{F}_2} \alpha \, \mathrm{d}\mathcal{F} = \alpha_i \cdot (\mathcal{F}_2 - \mathcal{F}_1),
\end{equation}
where $\mathcal{F}_1$ and $\mathcal{F}_2$ are the slopes of the two branches at the self-intersection point, and $\alpha_i$ is the corresponding celestial coordinate. 
So for rotating wormholes, the two areas enclosed on either side of the self-intersection point are equal, as illustrated in Fig. (\ref{fig:WH-area}).
It is worth noticing that while the unstable circular orbits and the wormhole throat orbits approach each other in the $(\alpha,\beta)$ plane with the same slope, the corresponding curve in the $\alpha$-$\mathcal{F}$ diagram is continuous, yet not necessarily smooth.
  The equal-area law follows solely from the closedness of the swallowtail and also applies to wormhole shadows, which further confirms the universality of the equal-area law.

Given that $\mathcal{F}_1$ and $\mathcal{F}_2$ are associated with different branches, it is convenient to consider the order parameter solely on the unstable circular orbits.
This approach avoids complications arising from the wormhole throat orbit.
Following \eqref{eq:expand}, we directly define $\Delta\bmF=|\mathcal{F}-\mathcal{F}_c|$ along the unstable circular orbits.
It is straightforward to verify that $\Delta \bmF$ follows the scaling relations
\begin{equation}
\Delta \bmF \sim |\alpha - \alpha_c|^{1/2},\qquad\Delta \bmF \sim |\lambda - \lambda_c|^{1/2}.
\end{equation}
The result can be verified by numerical fitting.
One of the critical exponents is demonstrated in Fig. (\ref{fig:WH-ce}), where the slope in the log-log diagram is $0.4999$.
The critical exponents are identical to the exponents found in the KZ black hole and the running-$G$ Kerr black hole, providing strong evidence that the critical phenomenon associated with cusp formation is universal and independent of the specific metric function.

In summary, the shadow of a rotating traversable wormhole manifests all three universal features associated with cuspy shadows: a topological charge transition from $+1$ to $-1$, an equal-area law governing the self-intersection point, and a critical scaling with an exponent of $1/2$ near the onset of cusp formation. These properties arise purely from the global topology and local bifurcation structure of the shadow boundary, remaining entirely independent of the detailed spacetime metric. Their emergence in a horizonless compact object such as a wormhole compellingly demonstrates that such phenomena are not exclusive to black holes; rather, they constitute a universal fingerprint of strong-field gravity, applicable to any spacetime admitting multiple families of unstable circular orbits.

\section{Conclusion and discussion}
\label{con}

In this paper, we present a unified and systematic investigation of cusp formation in compact object shadows. We identify three universal features characterizing this phenomenon: a topological charge transition, an equal-area law, and a universal critical exponent. We demonstrate that these features apply universally across the KZ black hole, the running-$G$ Kerr black hole, and the rotating traversable wormhole.
For those geometries, the transition from a smooth quasi-circular shadow to a cuspy one is controlled by a critical parameter. 
At the critical point, the shadow boundary develops an inflection point in the $(\alpha,\mathcal{F})$ plane.
Beyond this threshold, the unstable circular orbits become self-intersecting, giving rise to a swallowtail structure and, consequently, cusps in the observable shadow.
We have shown that three phenomena universally accompany cusp formation
\begin{itemize}
  \item First, the topological charge defined via the Gauss-Bonnet theorem undergoes a discrete transition. For a smooth Jordan curve, the total turning angle equals $2\pi$, leading to $\delta = +1$. When a swallowtail structure forms, the total topological charge becomes $\delta = -1$. This flip from $\delta = +1$ to $\delta = -1$ signals a genuine topological phase transition in the shadow morphology, independent of the specific spacetime geometry.
  \item Second, the self-intersecting structure obeys an equal-area law in the $(\alpha,\mathcal{F})$ diagram. The equal-area law originates from the closedness of the swallowtail loop and does not rely on thermodynamic analogies or the detailed form of the metric functions. It provides a precise and model-independent method to determine the self-intersection point $(\alpha_i,\beta_i)$.
  \item Third, the onset of cusp formation exhibits a universal critical scaling behavior. By identifying $\Delta \mathcal{F}$ as the relevant order parameter, we find that the critical exponents associated with both the control parameter and $\alpha$ are exactly $1/2$. Importantly, this scaling behavior is entirely robust against the explicit physical nature of the control parameter, whether it be the deformation parameter $\varepsilon$ in the KZ black hole, the running coupling $G(z)$ in the modified Kerr spacetime, or the redshift parameter $\lambda$ in the wormhole case. The emergence of the exponent $1/2$ firmly places cusp formation within the mean-field universality class, revealing a profound underlying connection between black hole optics and the general theory of critical phenomena.
\end{itemize}
The rotating traversable wormhole serves as a nontrivial, horizonless example that further corroborates this universality. Although its shadow is governed by two distinct families of photon orbits, the same three hallmark features emerge. Our results demonstrate that these phenomena originate from the intrinsic geometry of unstable circular orbits and their bifurcation structure near a critical point, remaining strictly independent of the detailed spacetime metric. %Ultimately, the universality of these features places cusp formation on firm geometric and topological ground, establishing it as a robust, model-independent probe of strong-field gravity.
These results reveal an unexpected universality in black hole optics and provide new insights into the geometric structure of photon orbits beyond the Kerr paradigm.

%From an observational perspective, our results offer new diagnostics for testing deviations from the Kerr metric. 
%The topological charge provides a global and coordinate-invariant observable. 
%Moreover, the equal-area law and critical scaling relations offer quantitative tools for characterizing the onset of non-Kerr behavior.

Note that the purpose of the current paper is not to claim the discovery of entirely new properties of compact-object shadows, but to clarify the origin, physical meaning, and universality of these properties. 
The topological charge transition, equal-area relation, and critical exponent should be understood as physical and geometrical constraints on ideal shadow boundaries generated by photon-orbit dynamics. 
The fact that the resulting laws take a simple geometric form is a manifestation of universality, similar to the role of Maxwell-type equal-area constructions and mean-field critical exponents in thermodynamics and critical phenomena.
Once the cusp/swallowtail structure is identified, its relevance to black hole shadows lies in the fact that the shadow boundary is not an arbitrary plane curve, but the image of a physical map from unstable null geodesics to the celestial coordinates of a distant observer.

From the perspective of ideal shadow morphology, our results provide theoretical indicators for characterizing possible deviations from the Kerr shadow. 
The topological charge captures the global structure of the ideal boundary, while the equal-area law and critical scaling relations quantify the self-intersection and the onset of cusp formation. 
It is worth emphasizing that the direct observation of cusps or the extraction of the associated topological quantities from current images would be highly challenging. These quantities should not be interpreted as directly extractable observables from current black hole images, but as geometric descriptors of the photon-orbit boundary and as theoretical input for future ray-tracing and phenomenological studies.
Moreover, the compact objects considered in this work are used as theoretical laboratories for studying possible non-Kerr shadow morphologies, rather than as phenomenologically established alternatives to the Kerr black hole. We do not claim that any particular exotic compact object or modified-gravity model is favored by current observations. 
Rather, our goal is to identify the universal geometric structure associated with cusp formation. From this perspective, the KZ non-Kerr black hole, the Kerr black hole with a running Newton coupling, and the rotating traversable wormhole provide explicit examples realizing the same cusp/swallowtail universality class. Thus, those properties should be understood as a theoretical classification of ideal shadow morphologies, not as evidence for any specific exotic object.

Several directions deserve further investigation. It would be interesting to explore whether similar universal structures arise in more general metrics with cuspy shadow, like \cite{Cunha:2017, Qian:2022}.
Further study may help clarify whether the universality class identified here is truly exhaustive.
The stability of cuspy features also warrants further study.

%In summary, we have demonstrated that the cuspy behavior of compact object shadows is governed by universal geometric principles. The topological charge flip, the equal-area law, and the critical exponent $1/2$ together form a coherent and model-independent framework for understanding cusp formation. 

%%%%%%%%%%%%%%%%%%%%%%%%%%%%%%%%%%%%%%%%%%%%%%%%%%%%%%%%%%%%%%%%%%%%%%%%%%%%%%%%%%%%%%%%%%%%%%%%%%%%

\section*{Acknowledgements}
We thank Wentao Liu and Shao-Wen Wei for the helpful discussions. 
This work is supported by the National Natural Science Foundation of China (Grant No. 12405073, No. 12305065, No. 12547103 and No. 12247101) and the Natural Science Foundation of Tianjin (Grant No. 25JCQNJC01920). PC is partially supported by Tianjin University Self-Innovation Fund Extreme Basic Research Project (Grant No. 2025XJ21-0007).

%%%%%%%%%%%%%%%%%%%%%%%%%%%%%%%%%%%%%%%%%%%%%%%%%%%%%%%%%%%%%%%%%%%%%%%%%%%%%%%%%%%%%%%%%%%%%%%%%%%%

%\newpage
%\appendix

%%%%%%%%%%%%%%%%%%%%%%%%%%%%%%%%%%%%%%%%%%%%%%%%%%%%%%%%%%%%%%%%%%%%%%%%%%%%%%%%%%%%%%%%%%%%%%%%%%%%

% \bibliography{ref}
% \bibliographystyle{utphys}

%\providecommand{\href}[2]{#2}
\begingroup\raggedright\endgroup

%%%%%%%%%%%%%%%%%%%%%%%%%%%%%%%%%%%%%%%%%%%%%%%%%%%%%%%%%%%%%%%%%%%%%%%%%%%%%%%%%%%%%%%%%%%%%%%%%%%%

\end{document}